\documentclass[useAMS,usenatbib,usegraphicx]{mn2e}
\usepackage{amsmath,amssymb,epsfig}
\usepackage{graphicx}
\usepackage{float}
\usepackage{breqn}
\usepackage{placeins}
\usepackage{hyperref}
\usepackage{natbib,ifthen}
\usepackage{bm}
\usepackage{xcolor}
\DeclareMathOperator\erf{erf}

\newcommand{\adsurl}[1]{\href{#1}{ADS}}

\newcommand\ba{\begin{eqnarray}}
\newcommand\ea{\end{eqnarray}}
\newcommand\be{\begin{equation}}
\newcommand\ee{\end{equation}}

\renewcommand\P{{\cal P}}

\newcommand\gsim{ \lower .75ex \hbox{$\sim$} \llap{\raise .27ex \hbox{$>$}} }
\newcommand\lsim{ \lower .75ex \hbox{$\sim$} \llap{\raise .27ex \hbox{$<$}} }

\newcommand{\Mpc}{\text{Mpc}}

\newcommand{\vk}{{\bm{k}}}

\newcommand{\vq}{{\bm{q}}}
\newcommand{\vr}{{\bm{r}}}

\newcommand{\grad}{{\nabla}}

\title[Gaussianizing transforms]{Perturbative Gaussianizing transforms for cosmological fields}

\author[Alex Hall and Alexander Mead]{\parbox\textwidth{Alex Hall$^{1}$\thanks{ahall@roe.ac.uk} and 
Alexander Mead${^{2,3}}$}\\\\
$^{1}$Institute for Astronomy, University of Edinburgh, Royal Observatory, Blackford Hill, Edinburgh, EH9 3HJ, U.K.\\
$^{2}$Department of Physics and Astronomy, University of British Columbia, 6224 Agricultural Road, Vancouver, BC V6T 1Z1, Canada\\
$^{3}$CITA National Fellow; Canadian Institute for Theoretical Astrophysics, University of Toronto, ON M5S 3H8, Canada}

\date{Accepted XXX. Received YYY; in original form ZZZ}

\pubyear{2017}

\begin{document}
\label{firstpage}
\pagerange{\pageref{firstpage}--\pageref{lastpage}}
\maketitle

\begin{abstract}
Constraints on cosmological parameters from large-scale structure have traditionally been obtained from two-point statistics. However, non-linear structure formation renders these statistics insufficient in capturing the full information content available, necessitating the measurement of higher-order moments to recover information which would otherwise be lost. We construct quantities based on non-linear and non-local transformations of weakly non-Gaussian fields that Gaussianize the full multivariate distribution at a given order in perturbation theory. Our approach does not require a model of the fields themselves and takes as input only the first few polyspectra, which could be modelled or measured from simulations or data, making our method particularly suited to observables lacking a robust perturbative description such as the weak-lensing shear. We apply our method to simulated density fields, finding a significantly reduced bispectrum and an enhanced correlation with the initial field. We demonstrate that our method reconstructs a large proportion of the linear baryon acoustic oscillations, improving the information content over the raw field by 35\%. We apply the transform to toy 21\,cm intensity maps, showing that our method still performs well in the presence of complications such as redshift-space distortions, beam smoothing, pixel noise, and foreground subtraction. We discuss how this method might provide a route to constructing a perturbative model of the fully non-Gaussian multivariate likelihood function.
\end{abstract}

\maketitle

\section{Introduction}
\label{sec:intro}

Planned cosmological surveys will produce an abundance of data which may be used to constrain cosmological models. In particular, surveys such as Euclid\footnote{\url{http://sci.esa.int/euclid/}}, WFIRST\footnote{\url{https://wfirst.gsfc.nasa.gov/}}, the Large Synoptic Survey Telescope\footnote{\url{https://www.lsst.org/}}, the Square Kilometre Array\footnote{\url{http://skatelescope.org/}}, and the Dark Energy Spectroscopic Instrument\footnote{\url{http://desi.lbl.gov/}} aim to make precise measurements of the properties of dark energy and other physics beyond the Standard Model, through a combination of galaxy number counts, weak gravitational lensing, and 21\,cm intensity mapping.

In order to fully realize the potential of these probes, it is important to ensure that as much information as possible is contained within statistics derived from the data. Measurements of large-scale structure are usually made with two-point statistics such as the power spectrum or correlation function, but as the observables are generally non-Gaussian distributed these quantities do not capture all the available information. Instead, non-linear gravitational collapse generates non-zero higher-order cumulants such as bispectra and trispectra~\citep{Peebles}, as well as distinct signatures in measures of topology such as the Minkowski functionals~\citep{1996app..conf..251P,2017MNRAS.467.3361W}. Ideally all these quantities should be measured by a survey to maximize the information content on the cosmological parameters, but the high dimensionality of the problem and the difficulty of modelling their covariances renders this an impractical task.

An alternative approach to capturing information contained in the higher-order cumulants is provided by Gaussianizing transforms. Typically, these involve either a local transform of the observable at each point using a logarithm or similar function~\citep{2009ApJ...698L..90N, 2011MNRAS.418..145J, 2013ApJ...763L..14M, 2017arXiv170402960L}, or remapping approaches which aim to undo non-linear evolution (e.g.~\citealt{2007ApJ...664..675E, 2016arXiv161109638Z} for galaxy surveys, \citealt{2016PhRvL.117o1102L} for cosmic microwave background lensing, and \citealt{2017JCAP...09..012O} for 21\,cm intensity mapping). Other approaches include iterative reconstruction~\citep{1992ApJ...391..443N, 1999MNRAS.308..763M, 2012MNRAS.425.2443K}, rank-order Gaussianization~\citep{1992MNRAS.254..315W}, clipping~\citep{2011PhRvL.107A1301S}, and wavelet transforms~\citep{2011ApJ...728...35Z, 2013MNRAS.436..759H}.

In this work we consider Gaussianization from a perturbative perspective by treating the first few non-Gaussian moments as small quantities. Whilst this will inevitably restrict the range of scales on which our approach is valid, it will give us a simple way to connect Gaussianization with perturbative models for the fields themselves, and provides a principled way to construct transforms without the use of ad-hoc functions. Our aim is to Gaussianize the full multivariate distribution of the observable, which necessitates the use of non-local transforms, i.e. transforms which mix the field at different spatial locations. We will see that this construction provides a generalization of the perturbative local transform of~\citet{2013MNRAS.434.2961C}. 

The ultimate goal of Gaussianization is to enhance the information content of power spectrum estimates by recovering information contained in the higher-order moments. A by-product of this is that the mean of these estimates should be more linear, since there is strong evidence that the initial conditions for large-scale structure are Gaussian. Thus, Gaussianization should both reduce error bars on power spectrum estimates and allow for more straightforward modelling of the signal\footnote{The caveat to this is that by focussing on cumulants we lose sensitivity to information not contained in the moment hierarchy - the lognormal distribution for example cannot be entirely expressed in terms of its cumulants~\citep{1991MNRAS.248....1C, 2011ApJ...738...86C}. However the density field is not exactly lognormal, and we restrict to large scales where the distribution is only weakly non-Gaussian.}.

In Section~\ref{sec:gauss} we present the mathematical formalism of our method and connect it cosmological fields. In Section~\ref{sec:dm} we apply our method to simulations of the dark matter density field and investigate its performance with a broad range of diagnostics. In Section~\ref{sec:21cm} we apply the method to toy realizations of 21\,cm maps, and we conclude in Section~\ref{sec:conc}. In Appendix~\ref{app:edge} we provide an alternative derivation of our transform based on the Edgeworth expansion.

\section{Gaussianizing transforms}
\label{sec:gauss}

\subsection{Polynomial expansion}
\label{subsec:gengauss}

In this section we derive general conditions on Gaussianizing transforms using a perturbative approach following~\citet{McCullagh}. Let $X^i$ be the $i^{th}$ element of a random vector $\bm{X}$ having some weakly non-Gaussian distribution. Without loss of generality, we subtract off the mean such that $\langle X^i \rangle = 0$ and normalize by the covariance such that $\langle X^i X^j \rangle = \delta^{ij}$, where angle brackets denote expectations over the distribution of $\bm{X}$ and $\delta^{ij}$ is the Kronecker delta. We then introduce an order-counting parameter $\epsilon$ such that $\langle X^i X^j X^k \rangle_c = \epsilon \kappa^{ijk}$ and $\langle X^i X^j X^k X^m \rangle_c = \epsilon^2 \kappa^{ijkm}$ and similarly for higher moments, where $\langle ... \rangle_c$ denotes the fully-connected part or cumulant. The definition of weak non-Gaussianity is then that $\epsilon \ll 1$. We denote with the notation $\mathcal{O}(n)$ a quantity whose leading order $\epsilon$-dependence is $\epsilon^n$, and by definition the objects $\kappa$ are $\mathcal{O}(1)$ such that as $\epsilon$ tends to zero $X^i$ is Gaussian. The $n^{th}$ cumulant of $X^i$ is then $\mathcal{O}(n-2)$. For example, if $X^i$ were the sum of $N$ statistically independent random vectors each having zero mean we would have $\epsilon \sim \sqrt{1/N}$, and if $X^{i}$ were a scalar field having quadratic local non-Gaussianity we would have $\epsilon \sim f_{\mathrm{NL}}$.

We seek a vector $Y^i(\bm{X})$ whose distribution is Gaussian at each order up to some given order in $\epsilon$. An obvious place to start would be to expand $Y^i(\bm{X})$ in a power series as
\begin{align}
&Y^i = X^i + \epsilon \left(m_{1,1}^i + m_{1,2}^{ij}X^j + m_{1,3}^{ijk}X^j X^k + ... \right) \nonumber \\ 
& + \epsilon^2 \left( m_{2,1}^i + m_{2,2}^{ij}X^j + m_{2,3}^{ijk}X^j X^k + m_{2,4}^{ijkm}X^jX^kX^m + ... \right) \nonumber \\
& + ...,
\label{eq:gtrans}
\end{align}
with repeated indices implicitly summed over. The problem is highly underconstrained since there are infinitely many choices for the number of non-zero unknown coefficients $m_{a,b}$. Once such a choice has been made however, we can derive constraints on the $m_{a,b}$ by requiring the non-Gaussian cumulants of $Y^i$ to vanish at each order. This requires that $m_{a,a+2}\neq 0$ for all $a \leq n$ when working to $\mathcal{O}(n)$. The simplest construction then enforces a triangular condition such that $m_{a,b} = 0$ for all $a \leq n$ and $b \geq a+3$ at $\mathcal{O}(n)$. At $\mathcal{O}(2)$ for example this would correspond to Equation~\eqref{eq:gtrans} with terms denoted by ellipses set to zero. Setting the third cumulant of $Y^i$ to zero at this order then requires that
\begin{align}
&m_{1,3}^{(ijk)} =-\frac{1}{6}\kappa^{ijk}, \nonumber \\
& m_{2,3}^{(ijk)} = -\frac{1}{2} m_{1,2}^{\left(i|r\right.}\kappa^{\left.|jk\right)r} - 2m_{1,2}^{\left( i | r \right.} m_{1,3}^{\left. |jk\right)r},
\label{eq:skewfix}
\end{align}
where round brackets denote total symmetrization on the enclosed indices and vertical lines denote indices excluded from the symmetrization. We can constrain $m_{2,4}^{ijkm}$ by requiring a vanishing fourth cumulant at leading order, which yields
\begin{equation}
 m_{2,4}^{(ijkm)} = -\frac{1}{24}\kappa^{ijkm} - m_{1,3}^{\left(ij|r\right.}\kappa^{\left.|km\right)r} -2 m_{1,3}^{\left(ij| r\right.}m_{1,3}^{\left. |km \right) r}.
\label{eq:kurtfix}
\end{equation}
Similarly $m_{1,1}^{i}$ and $m_{2,1}^i$ can be chosen to set the mean of $Y^i$ to some desired value at each order, while $m_{1,2}^{ij}$ and $m_{2,2}^{ij}$ are constrained by requiring the covariance of $Y^i$ to have some desired value at each order. Finally, $m_{2,3}^{ijk}$ can be constrained using Equation~\eqref{eq:skewfix} once $m_{1,2}^{ij}$ has been chosen. Note that when the dimensionality of $\bm{X}$ is greater than one the $m_{a,b}$ coefficients are still not uniquely determined, since only the totally symmetric parts are constrained in the above construction.

A simple choice for the coefficients sets $m_{1,3}^{ijk}$ and $m_{2,3}^{ijk}$ equal to the right-hand sides of Equation~\eqref{eq:skewfix} and $m_{2,4}^{ijkm}$ to the right-hand side of Equation~\eqref{eq:kurtfix}. Fixing the constant and linear terms such that $Y^i$ has zero mean and unit diagonal covariance (they can be chosen to give our Gaussian variate a desired mean and covariance, as long as Equation~\ref{eq:skewfix} is satisfied) yields
\begin{align}
Y^i &= X^i - \frac{\epsilon}{6} \left(\kappa^{ijk}X^jX^k - \kappa^{irr}\right) \nonumber \\
& - \epsilon^2\left[\left(\frac{1}{24}\kappa^{ijkm} - \frac{1}{9}\kappa^{ijr}\kappa^{kmr}\right)X^jX^kX^m \right. \nonumber \\
& \left. - \left(\frac{1}{8}\kappa^{ijrr} - \frac{1}{12}\kappa^{irs}\kappa^{jrs} - \frac{1}{9}\kappa^{ijr}\kappa^{rss}\right)X^j\right].
\label{eq:gaussY}
\end{align}
In one dimension, this takes the simplified form
\begin{equation}
Y = X - \frac{\kappa_3}{6}(X^2 - 1) - \frac{\kappa_4}{24}(X^3 - 3X) + \frac{\kappa_3^2}{36}(4X^3 - 7X),
\label{eq:1Dgtrans}
\end{equation}
where $\kappa_3$ and $\kappa_4$ are the dimensionless skew and kurtosis respectively, which agrees with the expression in~\citet{McCullagh}. Up to irrelevant constant and linear terms, Equation~\eqref{eq:1Dgtrans} is identical to the square-root of the `optimal observable', $o(\delta)$, presented in~\citet{2013MNRAS.434.2961C} and shown to capture all the information at this order in perturbation theory. The equivalence is not surprising, since a zero-mean Gaussian variate may be described entirely in terms of its variance, for which the quantity $X^2$ is a sufficient statistic. Equation~\eqref{eq:gaussY} can be seen as a multivariate generalization of the \citet{2013MNRAS.434.2961C} observable.

In Appendix~\ref{app:edge} we provide an alternative derivation of Equation~\eqref{eq:1Dgtrans} based on the Edgeworth expansion.

\subsection{Application to cosmological fields}
\label{subsec:cosmofields}

The formalism presented so far is applicable to any weakly non-Gaussian field. Specializing now to a statistically homogeneous and isotropic field in Fourier space, $\delta(\vk)$, the standardized variate $X^i$ introduced in the previous section can be constructed as
\begin{equation}
X^i = \sqrt{\frac{2}{P(k)V}} \left[ \mathrm{Re}\, \delta(\vk_i) \Theta(\vk_i) + \mathrm{Im}\, \delta(\vk_i) \Theta(-\vk_i)\right],
\end{equation}
where $\Theta(\vk)$ is the Heaviside step function. Note that there are alternative constructions for $X^{i}$ depending on how one vectorizes the Fourier-space density field and the above choice is arbitrary amongst these alternatives, but there is no ambiguity in the final transform when written in terms of $\delta(\vk)$. We can now plug this variate into the expressions of the previous section to derive the necessary constraints for a Gaussianizing transform. For example, the cumulant $\kappa^{ijk}$ now becomes
\begin{align}
&\epsilon \kappa^{ijk} = \frac{1}{\sqrt{2V P(k_i)P(k_j)P(k_k)}}\left\{B(\vk_i,\vk_j,\vk_k)\delta_{\vk_i+\vk_j,\vk_k} \right. \nonumber \\ 
&\times \Theta(\vk_i)\Theta(\vk_j)\Theta(\vk_k) + \left[B(\vk_i,-\vk_j,\vk_k)\delta_{\vk_j-\vk_i,\vk_k} \right. \nonumber \\
& \left. + B(\vk_i,\vk_j,-\vk_k)\delta_{\vk_j+\vk_i,\vk_k} - B(\vk_i,\vk_j,\vk_k)\delta_{\vk_j+\vk_i,-\vk_k}\right] \nonumber \\
& \left. \times \Theta(\vk_i)\Theta(-\vk_j)\Theta(-\vk_k)\right\} + \mathrm{perms.},
\end{align}
where $B$ is the bispectrum and $\mathrm{perms.}$ refers to the two additional cyclic permutations of the indices $\{i,j,k\}$\footnote{Our $\epsilon \kappa^{ijk}$ is equivalent to the normalized cumulant $p^{(3)}V^{-1/2}$ of~\citet{2007ApJS..170....1M}.}. Adopting the transformation of Eq~\eqref{eq:gaussY} gives
\begin{align}
&\tilde{\delta}(\bm{k}) = \delta(\bm{k}) - \frac{1}{6V} \sum_{\vq} \frac{B(\bm{k},\vq,-\bm{k}-\vq)}{P(q)P(|\bm{k} + \vq|)} \delta(\vq + \bm{k}) \delta(-\vq) \nonumber \\
&+ \frac{\delta(\vk)}{VP(k)}\sum_{\vq}\left[\frac{T(\vk,-\vk,\vq,-\vq)}{8P(q)} - \frac{B(\vk,\vq,-\vk-\vq)^2}{12P(q)P(|\vq + \vk|)} \right] \nonumber \\
& - \frac{1}{V^2}\sum_{\vq,\vr}\left[ \frac{T(\vk,\vq,\vr,-\vk-\vq-\vr)}{24} \right. \nonumber \\
&\left. - \frac{B(\vk,\vq,-\vk-\vq)B(\vk+\vq,\vr,-\vk-\vq-\vr)}{9P(|\vk+\vq|)}\right]\nonumber \\
& \times \frac{\delta(-\vq)\delta(\vr)\delta(\vq+\vr+\vk)}{P(q)P(r)P(|\vk+\vq+\vr|)},
\label{eq:wrongskew}
\end{align}
where $B$ is the bispectrum, $T$ is the trispectrum, and $P$ is the power spectrum\footnote{Formally $P(k)$ is the \emph{non-linear} power spectrum, but the difference is of higher order than Equation~\eqref{eq:wrongskew}.}. Equations~\eqref{eq:skewfix}, \eqref{eq:kurtfix} and \eqref{eq:wrongskew} are the main results of this work, and provide explicit formulae for Gaussianizing a weakly non-Gaussian field based only on knowledge of the first few cumulants or polyspectra. An expression similar to Equation~\eqref{eq:wrongskew} arises when performing the reverse construction, i.e. forming a non-Gaussian quantity from a Gaussian variable. Equation~\eqref{eq:wrongskew} essentially undoes such a construction, with the leading order term subtracted identical to that added to a Gaussian field to obtain a quantity having a given bispectrum (e.g., \citealt{2010JCAP...10..022W}).

The expression Equation~\eqref{eq:wrongskew} produces a Gaussianized field at order $\sim \mathcal{O}(T)\sim \mathcal{O}(B^2)$ for any weakly non-Gaussian field $\delta(\vk)$ satisfying statistical homogeneity and isotropy. This could be the real-space dark matter density field or 21\,cm brightness temperature on large spatial scales, or the cosmic shear field from weak gravitational lensing; Equation~\eqref{eq:wrongskew} is applicable to fields of any spatial dimension.

We note finally that the general formalism of this section may be used to Gaussianize observables for which a perturbative expansion is difficult to write down, such as the weak lensing convergence field. Using a numerical form for the first few cumulants derived from simulations, our method allows for non-Gaussian information to be moved to the power spectrum at leading order, which should improve on previous work that restricted to local transforms (e.g.~\citealt{2011MNRAS.418..145J}). One may worry that misestimation of these cumulants due to mismatches between the true cosmological parameters and those used in the simulations might degrade the performance of our Gaussianizing transform. We note that the leading order dependence on both the growth factor and $\sigma_8$ in the polyspectra cancels with that of the power spectra in the ratios of Equation~\eqref{eq:wrongskew}. Higher-order terms in the bispectrum and trispectrum bring further cosmology dependence which does not cancel with the power spectra, and although these should be suppressed in the perturbative regime in which we work, further tests will be necessary to precisely quantify the residual effects. The cumulants could also be measured from data, although high signal-to-noise measurements of all the configurations which dominate the sums in Equation~\eqref{eq:wrongskew} are required. It is important to note that these measurements should not come from the same density field used for the Gaussianizing transform, as the extra correlations introduced are not accounted for in our formalism.

\section{Application to the dark matter density field}
\label{sec:dm}

For a three dimensional field such as the dark matter density, the right-hand side of Equation~\eqref{eq:wrongskew} can be slow to compute since in general the large sums over wavevectors cannot be written as convolutions. This can be remedied however by making a different choice for the $m_{a,b}$ coefficients, suggested by standard Eulerian perturbation theory. Note that we have the freedom to do this as long as the conditions of Section~\ref{subsec:gengauss} are satisfied. At leading order the choice
\begin{equation}
m_{1,3}^{ijk} = -F_2^{(s)}(\bm{k}_j,\bm{k}_k) \delta_{\vk_j - \vk_i,\vk_k},
\label{eq:m13choice}
\end{equation}
where $F_2^{(s)}$ is the symmetrized second-order kernel of standard perturbation theory\footnote{This is given by $F_2^{(s)}(\vk_1,\vk_2) = \frac{5}{7} + \frac{1}{2}\hat{\vk}_1 \cdot \hat{\vk}_2 (\frac{k_1}{k_2} + \frac{k_2}{k_1}) + \frac{2}{7}(\hat{\vk}_1 \cdot \hat{\vk}_2)^2$, see e.g.~\citet{2002PhR...367....1B}.}, satisfies Equation~\eqref{eq:skewfix} when the bispectrum is replaced with its tree-level Einstein-de Sitter (EdS) form\footnote{Note that this form for the bispectrum only assumes EdS in the mode-coupling kernels, for which the cosmology dependence is weak~\citep{2002PhR...367....1B}.}. The Gaussianized density field at leading order (and in the continuum limit) is then
\begin{equation}
\tilde{\delta}(\vk) = \delta(\vk) - \int \frac{\mathrm{d}^3 \vq}{(2\pi)^3} F_2^{(s)}(\vq,\vk - \vq) \delta(\vq) \delta(\vk - \vq) - ...,
\end{equation}
which just amounts to subtracting off the leading-order non-linear part of the EdS density field. This will remove the leading order bispectrum, which is given by the expectation of products of linear terms with second-order terms.

The above argument suggests a Gaussianizing transform for the real-space dark matter density field is more straightforwardly derived by first writing down a perturbative expansion of the non-linear field and then inverting it order-by-order (see~\citealt{2012MNRAS.425.2443K,2017PhRvD..96b3505S} for applications of this using iterative approaches). Note that this approach runs against the main motivation for deriving Equation~\eqref{eq:wrongskew} since it requires a perturbative model for the field, which is not available in general. We chose this approach as it speeds up the computation of the transforms, but we note that for a two-dimensional field such as cosmic shear the sums in Equation~\eqref{eq:wrongskew} should be much quicker to compute.

Since we are interested in linearizing cosmological power spectra, it is necessary to work to third-order in the density field in order to remove both the leading-order loop power spectra ($P_{22}$ and $P_{13}$) and the leading-order non-Gaussian contribution to the power spectrum variance (the tree-level trispectrum). To third-order and assuming an EdS background, the expression to be inverted is
\begin{align}
\delta(\vk) &= \delta_L(\vk) + \int \frac{\mathrm{d}^3 \vq_1}{(2\pi)^3} F_2^{(s)}(\vq_1,\vk - \vq_1) \delta_L(\vq_1) \delta_L(\vk - \vq_1)  \nonumber \\
&+ \int \frac{\mathrm{d}^3 \vq_1}{(2\pi)^3}  \frac{\mathrm{d}^3 \vq_2}{(2\pi)^3} F_3^{(s)}(\vq_1,\vq_2,\vk - \vq_1-\vq_2) \nonumber \\
& \times \delta_L(\vq_1) \delta_L(\vq_2) \delta_L(\vk - \vq_1 - \vq_2),
\label{eq:series}
\end{align}
where $\delta_L$ is the linear field, and an explicit expression for $F_3^{(s)}$ may be found in~\citet{2002PhR...367....1B}. Fourier transforming this gives the corresponding expression in real space $\delta(\vr) = \delta_L(\vr) + \delta^{(2)}(\vr) + \delta^{(3)}(\vr) + ...$, with the second-order term given by
\begin{equation}
\delta^{(2)}[\delta_L] = \frac{17}{21}\delta_L^2 - \bm{\Psi}_L \cdot \grad \delta_L + \frac{2}{7} K_{L,ij}K_{L,ij},
\end{equation}
where the linear displacement field in Fourier space is $\bm{\Psi}_L(\vk) = i\frac{\vk}{k^2} \delta_L(\vk)$ and the linear tidal tensor is given by $K_{L,ij}(\vk) = \left(\frac{k_ik_j}{k^2} - \frac{1}{3}\delta_{ij}\right)\delta_L(\vk)$. The third-order part is given by
\begin{align}
& \delta^{(3)}[\delta_L] = \frac{341}{567} \delta_L^3 + \frac{11}{21} \delta_L K_{L,ij}K_{L,ij}  - \frac{4}{9}K_{L,ij} \grad_j \Psi_i^{(2)} \nonumber \\
& + \frac{2}{9}K_{L,ij}K_{L,jk}K_{L,ik} + \frac{1}{2} \Psi_{L,i}\Psi_{L,j} \grad_i \grad_j \delta_L  - \Psi_i^{(2)} \grad_i \delta_L \nonumber \\
& - \Psi_{L,i}K_{L,ij}\grad_j \delta_L - \frac{41}{21}\delta_L \Psi_{L,i} \grad_i \delta_L - \frac{4}{7} K_{L,jk}\Psi_{L,i} \grad_i K_{L,jk},
\label{eq:3rdorder}
\end{align}
where the second-order displacement field is given in Fourier space by\footnote{Note that $\vq$ is a wavevector and should not be confused with the initial spatial position vector of fluid elements.}
\begin{align}
  \bm{\Psi}^{(2)}(\vk) &= i \frac{3}{14} \frac{\vk}{k^2} \mathrm{F.T.} \left[\delta_L(\vr)^2 \vphantom{\frac12}\right. \nonumber \\
 & - \left. \mathrm{F.T.}^{-1}\left( \frac{q_i q_j}{q^2}\delta_L(\vq) \right)\mathrm{F.T.}^{-1}\left( \frac{q_i q_j}{q^2}\delta_L(\vq) \right) \right].
\label{eq:disp2}
\end{align}
In Equation~\eqref{eq:disp2}, $\mathrm{F.T.}$ denotes the Fourier transform and $\mathrm{F.T.}^{-1}$ its inverse.

Inverting the series expansion Equation~\eqref{eq:series} to third-order then yields an approximate expression for the linear density field
\begin{equation}
\delta_L \approx \delta - \delta^{(2)}\left[\delta - \delta^{(2)}[\delta]\right] - \delta^{(3)}[\delta],
\label{eq:HM}
\end{equation}
where $\delta^{(3)}[\delta]$ denotes the third-order density field Equation~\eqref{eq:3rdorder} with the non-linear field used in place of the linear field, and similarly for $\delta^{(2)}[\delta]$. Note that this is a non-local transform since the functionals $\delta^{(2)}[\cdot]$ and $\delta^{(3)}[\cdot]$ mix up different spatial locations\footnote{Keeping only second-order terms in Equation~\eqref{eq:HM} yields a transform equivalent to the `EF2' reconstruction of~\citet{2015PhRvD..92l3522S}.}. To enhance the convergence of this transform we could first derive an estimate for the linear field and then recompute the second and third-order terms in Equation~\eqref{eq:3rdorder} iteratively until convergence was reached~\citep{1999MNRAS.308..763M, 2012MNRAS.425.2443K}. We choose not to do this however in order to speed up the algorithm.

At leading order the field given by Equation~\eqref{eq:HM} should have a non-linear power spectrum of order $~(P_L/V)^3$, a bispectrum of order $~(P_L/V)^3$, and a trispectrum of order $~(P_L/V)^4$. However, due to the mode-coupling integrals in Equation~\eqref{eq:series} and the non-perturbative nature of $\delta(\vk)$ at large $k$, the non-linear fields entering the right-hand side of Equation~\eqref{eq:HM} have to be smoothed to suppress the contribution from small scales and ensure that the tree-level expressions for the cumulants dominate over their loop corrections. Additional smoothing will occur if the density field is observed at finite resolution. This inevitably worsens the accuracy of Equation~\eqref{eq:HM}, since the high-$k$ modes of the non-linear density contain contributions from the linear density at a range of scales dictated by mode-coupling. Smoothing the non-linear field on small scales destroys information about the linear field at much larger scales, therefore we do not expect the suppression of the non-linear power and non-Gaussian cumulants to be as good as the $P_L/V$ scaling indicated above.

Using the more general formalism of Section~\ref{subsec:gengauss} we could improve convergence of the transform with an expression for the bispectrum more accurate than the tree-level form which led to Equation~\eqref{eq:m13choice}. This more general formalism only requires that the dark matter density field is weakly non-Gaussian, in the precise sense that its cumulants become smaller at each order. It may well be the case that this property holds at smaller scales than the regime in which perturbation theory expressions for these cumulants are accurate. If this were the case, numerical expressions for the bispectrum and trispectrum derived from simulations could be used in our formalism across this range of scales.

\subsection{Tests on simulations}
\label{subsec:results}

In order to test our linearization expression Equation~\eqref{eq:HM}, we run a set of 40 dark-matter-only simulations differing only via the random numbers used to generate mode phases and amplitudes. The simulations use $512^3$ particles and were run using the \textsc{gadget-2} code \citep{2005MNRAS.364.1105S} in TreePM mode. The simulations take place in cubes of side length $1000\,h^{-1}\mathrm{Mpc}$ with gravitational softening set at $39\,h^{-1}\mathrm{kpc}$. The cosmological parameters were set to $\Omega_{\mathrm{c}}=0.25$, $\Omega_{\mathrm{b}}=0.05$, $\Omega_{\mathrm{v}}=0.7$, $h=0.7$, $\sigma_8=0.8$ and $n_{\mathrm{s}}=0.96$; neutrinos are considered massless and the effect of radiation on the expansion is ignored. Initial conditions were generated using the \textsc{N-genIC} package at an initial redshift of $99$ using a matter power spectrum appropriate for this cosmology from \textsc{CAMB} \citep{Lewis2000}. The simulations make no distinction between baryonic and cold-dark matter. Density fields were generated using cloud-in-cell interpolation of particles onto a $256^3$ grid. In this section we confine ourselves to the real-space dark matter density field at $z=1$.

\begin{figure}
\centering
\includegraphics[width=\columnwidth]{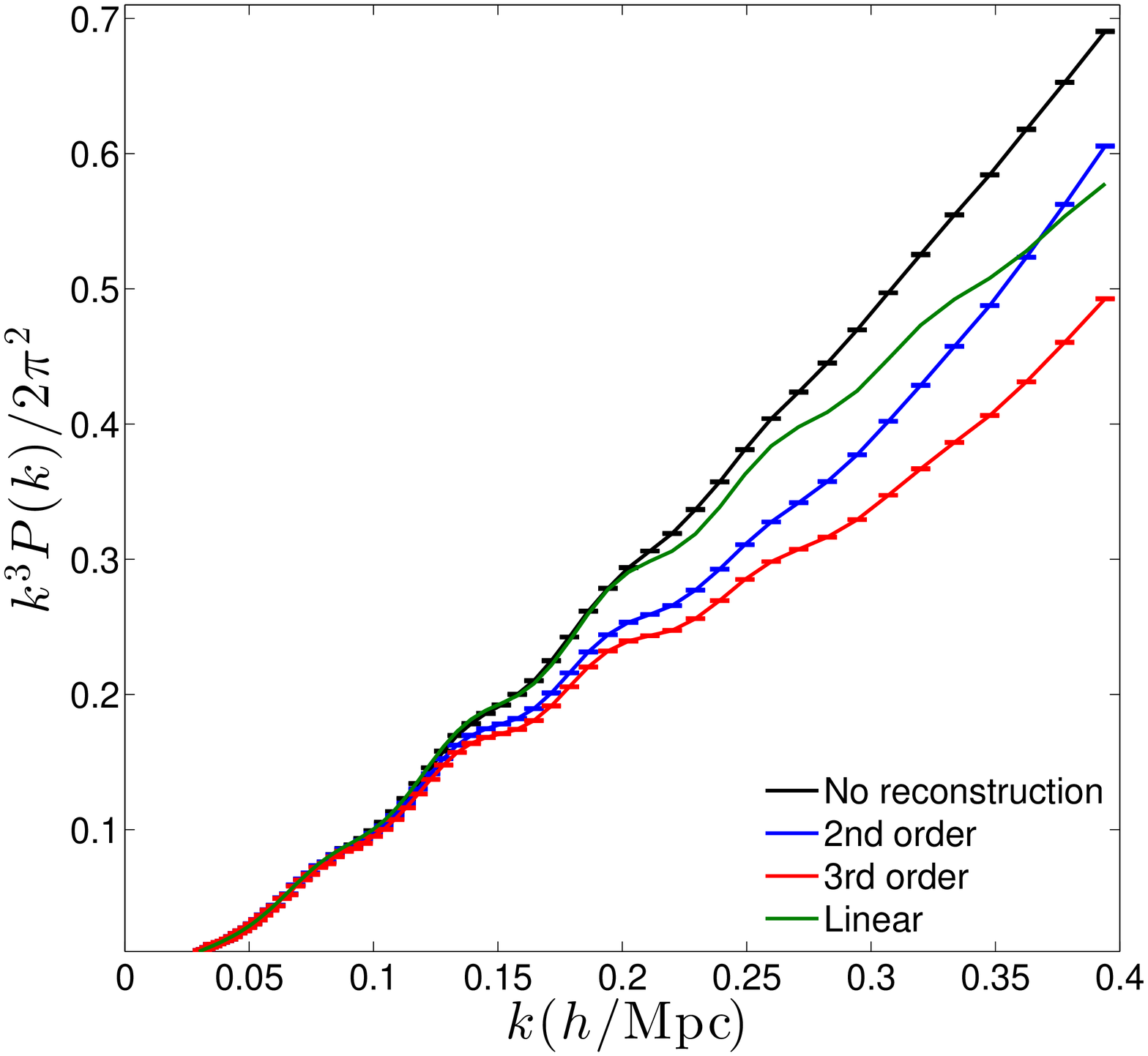}
\includegraphics[width=\columnwidth]{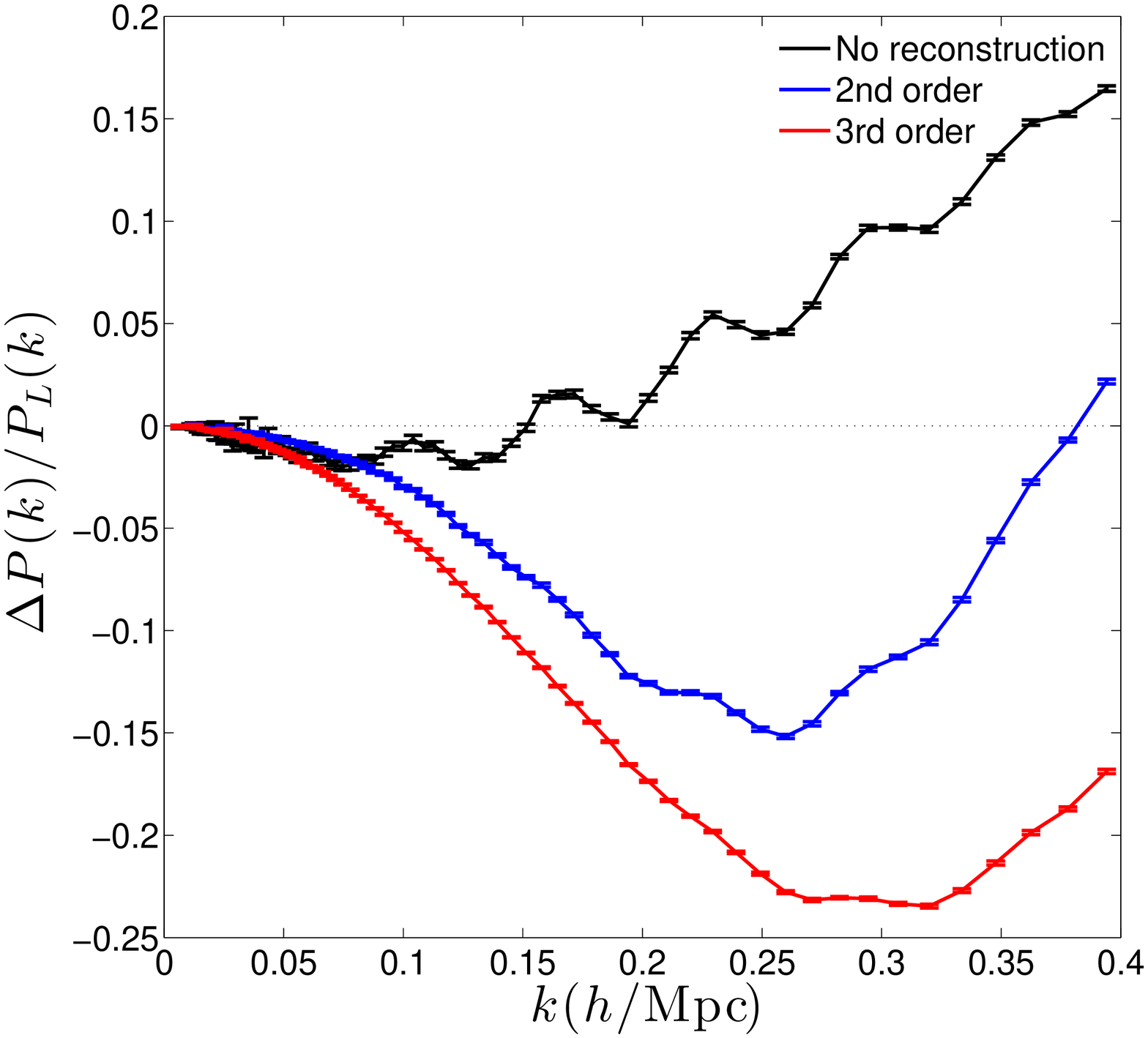}
\caption{\emph{Upper panel}: Average power spectrum from the simulations at $z=1$ for the unmodified field (black, upper points at $k=0.4 \, h^{-1} \Mpc$), the field corrected for second-order terms (blue, middle points at $k=0.4 \, h^{-1} \Mpc$), the field corrected for third-order terms (red, lower points at $k=0.4 \, h^{-1} \Mpc$) and the linear power spectrum at $z=1$ from \textsc{CAMB} (green solid). \emph{Lower panel}: Fractional differences of the power spectra from the linear power spectrum. Error bars in both panels are derived from the standard deviations of the 40 simulations.}
\label{fig:broadpdiff}
\end{figure}
 
We construct the right-hand side of Equation~\eqref{eq:HM} using the FFTW algorithm to compute the required non-linear combinations of the density field for each simulation realization. The density fields entering into these non-linear terms are first smoothed isotropically as $\delta(\vk) \rightarrow S(k)\delta(\vk)$. We experimented with different functional forms for the kernel $S(k)$, and found that a Gaussian filter $S(k) = \exp(-k^2/k_s^2)$ gave the best results. We test Equation~\eqref{eq:HM} using both the full third-order expression and a form valid at second-order given by $\delta_L \approx \delta - \delta^{(2)}[\delta]$. The inverse smoothing scale $k_s$ must be chosen to be small enough that non-perturbative modes of $\delta(\vk)$ do not invalidate our perturbative approach, but large enough that our expression has a non-zero impact on the density field.  We experimented with different choices of $k_s$ and found that best results were obtained with $k_s = 0.2 \, h\, \mathrm{Mpc}^{-1}$ when constructing the $\delta^{(2)}[\delta]$ part of the second term on the right-hand side of Equation~\eqref{eq:HM}, and $k_s = 0.3 \, h\, \mathrm{Mpc}^{-1}$ in the third-order term and the $\delta$ part of the second term on the right-hand side of Equation~\eqref{eq:HM}. We found the results to be fairly sensitive to these choices, but improved performance over the unmodified field was obtained in the Gaussianity tests for all choices of smoothing scales and kernels with which we experimented. We plot the results of this section to a maximum wavenumber of $k_{\rm{Ny}}/2 \approx 0.4 \, h \, \rm{Mpc}^{-1}$ to mitigate the influence of aliasing on small scales.

\subsubsection{Power spectrum}
\label{subsubsec:band}

In Fig.~\ref{fig:broadpdiff} we plot the mean power spectra of the second and third-order transformed fields in the upper panel, and the fractional differences of these with respect to the power spectrum of the initial conditions in the lower panel. Error bars are obtained from the standard deviation of the 40 simulation realizations. As this figure shows, the correction terms of Equation~\eqref{eq:HM} perform quite poorly in linearizing the power spectrum, with both the second-order and third-order corrections producing power spectra suppressed by tens of percent compared to the linear $P(k)$. The second-order scheme performs slightly better than the third-order transform, even showing improvement over the unmodified field on the largest scales. That our simple perturbative corrections fail to linearize the broadband power is probably a consequence of the breakdown of perturbation theory in describing the power spectrum in the quasi-linear regime (see, e.g.~\citealt{2009PhRvD..80d3531C}) and the loss of high-$k$ modes which contain linear power required at larger scales.

The residual bias will have to be modelled in order to prevent biases in the inferred cosmological parameters. We note however that modelling this bias with simulations comes at no extra computational expense, since simulations will have to be run anyway to model the non-linear field and its covariance, and our Gaussianizing transform is quick to compute. In contrast, information gain could only be otherwise achieved by measuring the higher-order polyspectra or increasing the survey volume, both of which are computationally and financially expensive. Whether our transforms really do increase the signal-to-noise of power spectrum estimates will be assessed below when we compute the covariance matrix.

We note finally that our method is applicable not only to the dark matter density field but any weakly non-Gaussian cosmological field, which may not generally have well-understood physical properties.

\subsubsection{Bispectrum and Trispectrum}
\label{subsubsec:bispec}
%

\begin{figure}
\centering
\includegraphics[width=\columnwidth]{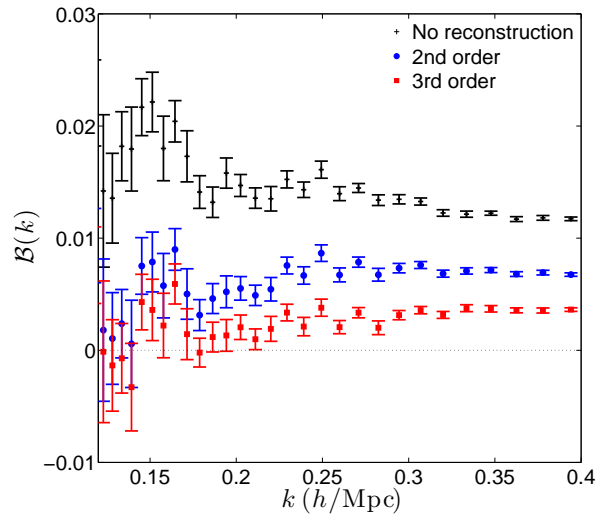}
\caption{Average normalized folded bispectrum from the simulations at $z=1$ for the unmodified field (black crosses), the field corrected for second-order terms (blue circles) and the field corrected for third-order terms (red squares). We only plot the highest few $k$-bins here for clarity, as the signal-to-noise on the measurements drops to zero at lower $k$.}
\label{fig:bfolded}
\end{figure}

A Gaussian field has zero bispectrum, and so as a first test of Equation~\eqref{eq:HM} we plot in Fig.~\ref{fig:bfolded} the normalized folded-shape bispectrum of the density field measured from simulations. The unnormalized folded bispectrum $B(k)$ is defined as $B(\vk_1,\vk_2,\vk_3)$ with $\vk_1 = \vk_2 = -2\vk_3 \equiv \vk$, and at leading-order (tree-level) this is given by the three-point function of the second-order field $\delta^{(2)}$ with two linear fields\footnote{In practice we allow the angle between $\vk_1$ and $\vk_3$ to vary by $\pm 20^{\circ}$ to boost the signal-to-noise on the measurement.}. Since we effectively subtract off the second-order term the resulting bispectrum should be proportional to loop integral terms of the form $\langle \delta_L \delta_L \delta^{(4)} \rangle \sim (P_L/V)^3$. We can define the normalized bispectrum as $\mathcal{B}(\vk_1,\vk_2,\vk_3) \equiv B(\vk_1,\vk_2,\vk_3)/\sqrt{VP(k_1)P(k_2)P(k_3)}$, which has the advantage of being invariant under the Fourier-space scaling $\delta(\vk) \rightarrow W(k)\delta(\vk)$. Reduction of $\mathcal{B}$ is a better measure of Gaussianity than $B$ since smoothing with $W(k)$ does not change the information content of the field and represents an uninteresting `trivial' Gaussianizing transform.

%
%
The results plotted in Fig.~\ref{fig:bfolded} are encouraging. The bispectrum has been significantly reduced by the Gaussianizing transforms across all scales where we have enough simulations to enable firm conclusions to be drawn, with a second-order transform resulting in a 40\% reduction and a third-order transform in a 70\% reduction. The unnormalized bispectrum is reduced by an order of magnitude by the transforms, comparable to the performance of clipping~\citep{2011PhRvL.107A1301S}, although this is largely due to an effective smoothing of the field.
%

\begin{figure}
\centering
\includegraphics[width=\columnwidth]{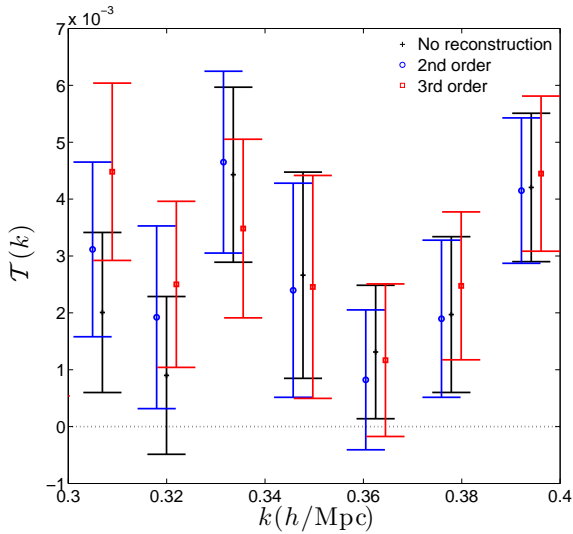}
\caption{Average normalized degenerate trispectrum from the simulations at $z=1$ for the unmodified field (black crosses), the field corrected for second-order terms (blue circles) and the field corrected for third-order terms (red squares). We only plot the highest few $k$-bins here for clarity, as the signal-to-noise on the measurements drops to zero at lower $k$.}
\label{fig:tdeg}
\end{figure}

In Fig.~\ref{fig:tdeg} we plot the normalized degenerate-shape trispectrum of the dark matter density field measured from simulations for the second and third-order corrections of Equation~\eqref{eq:HM}, and for the highest few $k$-bins where the signal-to-noise is largest. The degenerate shape corresponds to a trispectrum $T(\vk_1,\vk_2,\vk_3,\vk_4)$ with $\vk_1 = \vk_2 = -\vk_3 = -\vk_4 \equiv \vk$. Due to isotropy this depends only on the magnitude of the wavevector, and can be computed simultaneously with the power spectrum at no extra computational cost. Furthermore since the non-Gaussian contribution to the power spectrum variance is given by the trispectrum, it is of pivotal interest in testing the capability of Gaussianizing transforms to reduce this quantity. We define the normalized trispectrum $\mathcal{T}(\vk_1,\vk_2,\vk_3,\vk_4) \equiv T(\vk_1,\vk_2,\vk_3,\vk_4)/\sqrt{V^2P(k_1)P(k_2)P(k_3)P(k_4)}$, which is invariant under a Fourier-space scaling of the density field.

%
%

The noise on the trispectrum measurements in Fig.~\ref{fig:tdeg} is large due to the finite number of simulation realizations, and we see no significant difference between the different transformed fields. In contrast, we observe reductions of roughly 30\% and 45\% in the unnormalized trispectrum of the second-order and third-order transformed fields respectively. Most of this reduction comes from an effective smoothing of the density field, since there is no evidence of suppression in the normalized trispectrum plotted in Fig.~\ref{fig:tdeg}.

\subsubsection{Correlation with initial field}
\label{subsubsec:prop}

We have seen that our Gaussianizing transforms reduce the bispectrum of the non-linearly evolved density field, which hints at increased Gaussianity. However, it is unclear whether this is really due to the field behaving more like the initial Gaussian field or due to some other effect such as enhanced noise or the flow of information to higher moments. Furthermore, it is unclear how to interpret any increase in information contained in the power spectrum estimates that our method may enable (see~\citealt{2013MNRAS.436..759H} for a detailed discussion of this point). 

To elucidate the nature of the transformed fields, we construct the dimensionless cross-correlation coefficient $r(k)$ between the transformed field $\tilde{\delta}(\vk)$ and the initial field $\delta_L(\vk)$, given by
\begin{equation}
r(k) \equiv \frac{\langle \delta_L(\vk) \tilde{\delta}^*(\vk) \rangle}{\sqrt{\langle \delta_L(\vk) \delta_L^*(\vk) \rangle \langle \tilde{\delta}(\vk) \tilde{\delta}^*(\vk) \rangle}}.
\end{equation}
For each simulation realization we use the same initial field which was evolved under gravity to produce the density fields at $z=1$, which removes most of the cosmic variance in $r(k)$. We average the correlation coefficient over the 40 simulations to reduce this variance even further.

\begin{figure}
\centering
\includegraphics[width=\columnwidth]{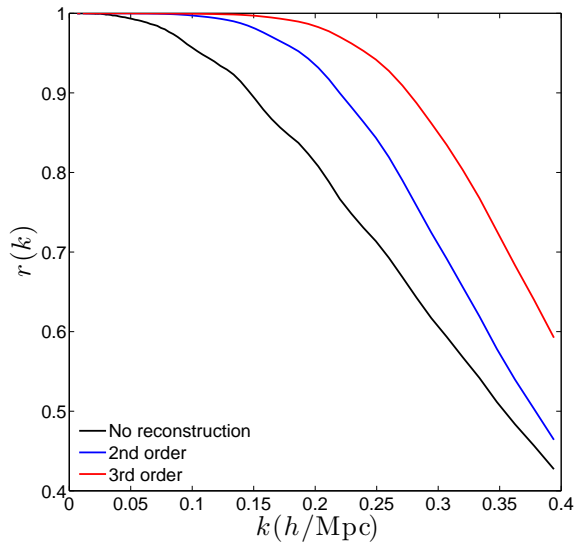}
\caption{Dimensionless cross-correlation coefficient of the initial Gaussian field with the unreconstructed field (black, lower curve), the second-order reconstructed field (blue, middle curve) and the third-order reconstructed field (red, upper curve). Note that the leading order cosmic variance cancels in this ratio.}
\label{fig:prop}
\end{figure}

In Fig.~\ref{fig:prop} we plot correlation coefficient for each of our Gaussianizing transforms. On large scales $r(k)$ tends to unity since the field is still in the linear regime and there is no mode-coupling, whilst on small scales the correlation drops to zero as mode-coupling from non-linear structure formation dominates any residual coherence with the initial field at a given wavenumber. Encouragingly, our transforms enhance the correlation with the initial field, with a third-order transform outperforming a second-order transform across all scales. This indicates that the Gaussianizing behaviour of our method observed in the normalized bispectrum is coming from an enhanced correlation of the density fields with the initial linear field.

\subsubsection{One-point probability density function}
\label{subsubsec:pdf}

As a further test of the Gaussianizing transform Equation~\eqref{eq:HM}, we construct the one-point probability density function (p.d.f.) of the unmodified, second-order transformed and third-order transformed density field at $z=1$. Since the density field is constructed on a coarse grid with a Nyquist frequency of $k_{\mathrm{Ny}} \sim 0.8 \, h \, \Mpc^{-1}$, the density field has been effectively smoothed on this scale prior to the construction of the p.d.f. We also constructed distributions after smoothing the unmodified and transformed fields on a scale $k_s = 0.2 \, h \, \Mpc^{-1}$ with a Gaussian filter. We refer to these two smoothing scales as the high-$k_s$ and low-$k_s$ respectively\footnote{Note that these smoothing scales should not be confused with the scales on which the density field is smoothed prior to the transforms being applied, since the operations do not commute.}.

\begin{figure*}
\centering
\includegraphics[width=\columnwidth]{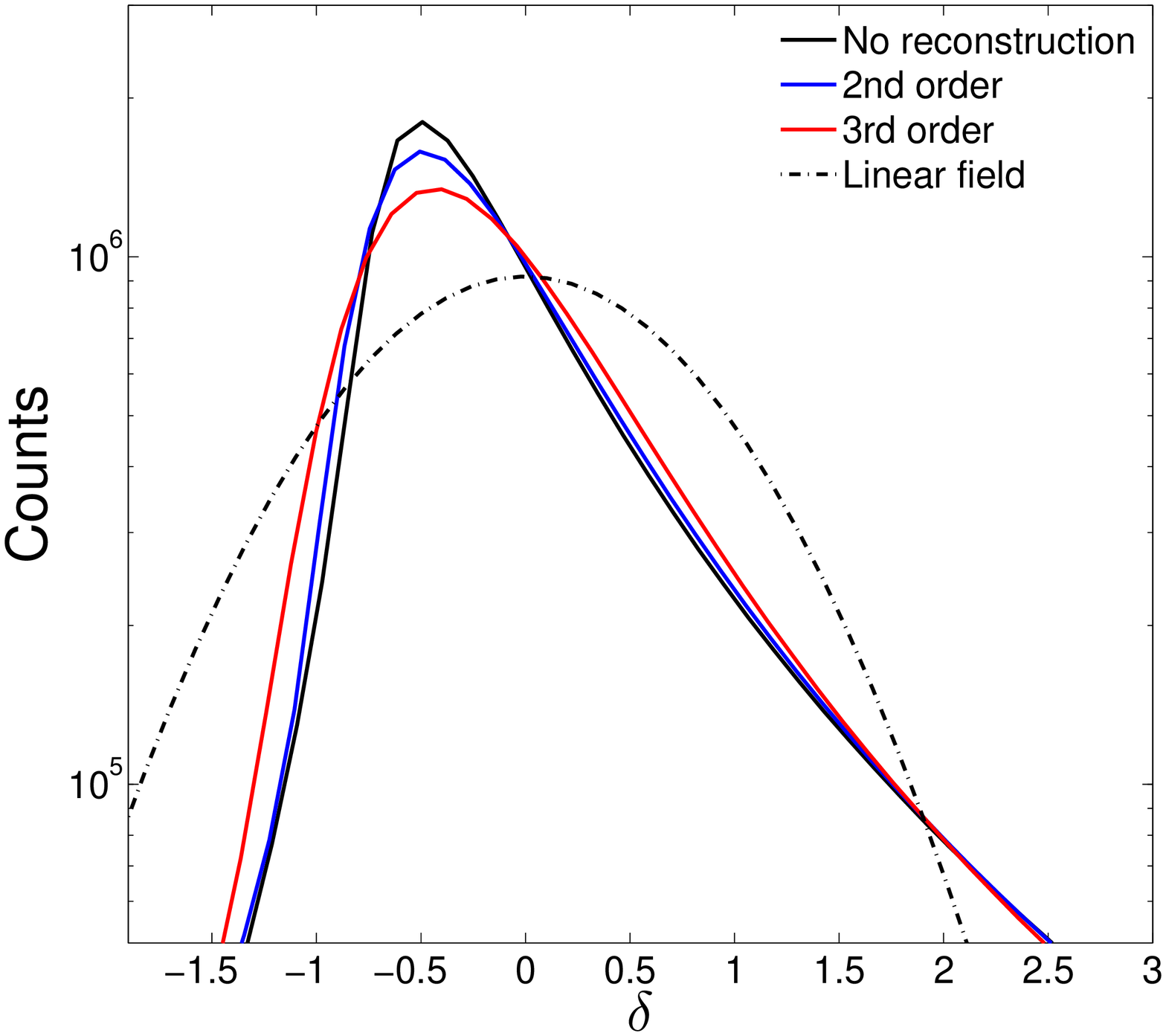}
\includegraphics[width=\columnwidth]{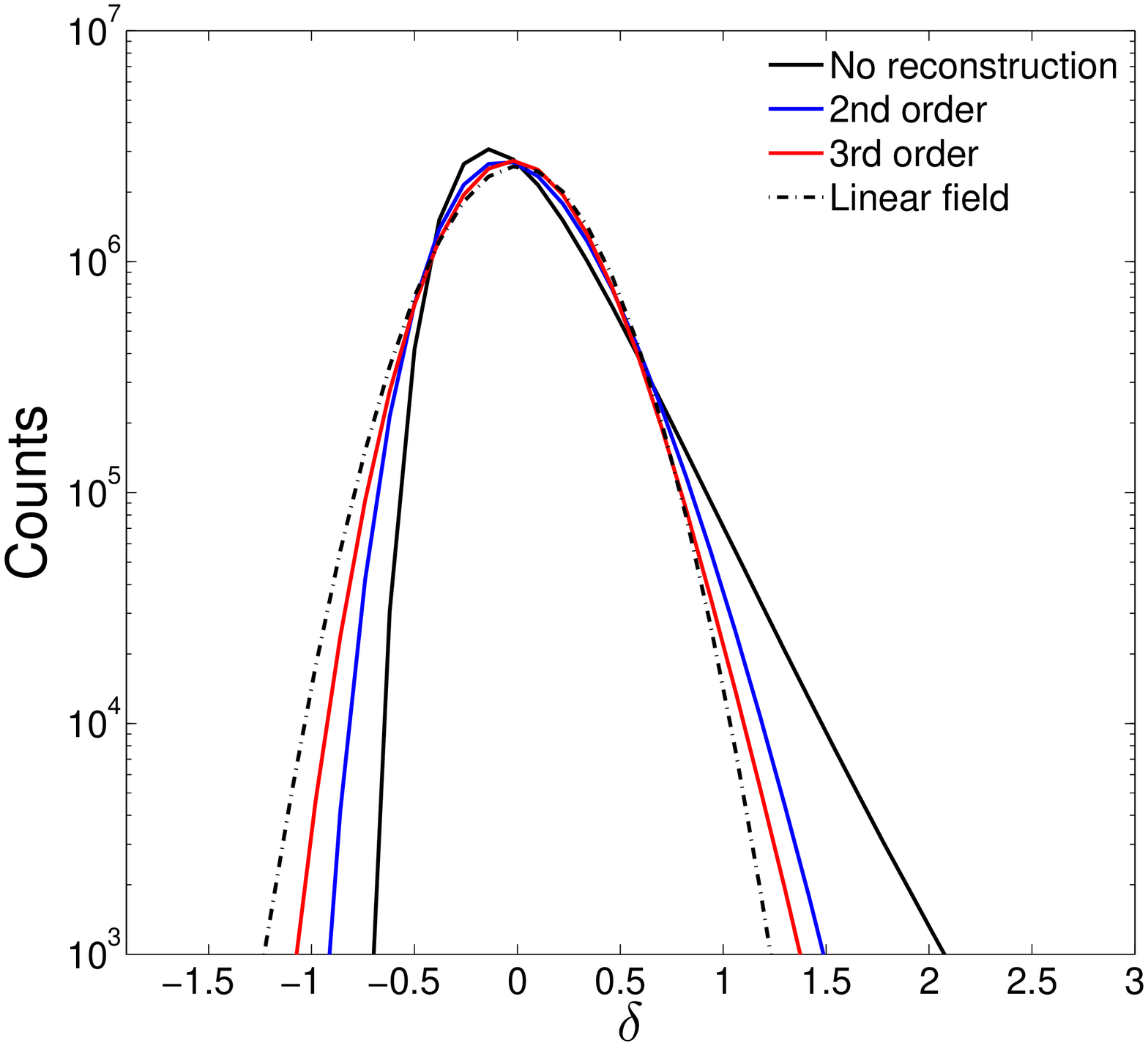}
\caption{\emph{Left panel}: One-point unnormalized p.d.f. of the unmodified density field smoothed on a scale $k_s = 0.8 \, h/\rm{Mpc}$ (black solid, upper curve at $\delta = -0.3$), the second-order transformed field (blue solid, middle curve at $\delta = -0.3$), the third-order transformed field (red solid, lower curve at $\delta = -0.3$), and the initial conditions scaled to $z=1$ with the linear growth factor (black dot-dashed). Error bars are smaller than the thickness of the curves. \emph{Right panel}: Same as left panel but with $k_s = 0.2 \, h/\rm{Mpc}$. Note that the stronger smoothing of non-linear scales has made all the distributions more Gaussian.}
\label{fig:pdfs}
\end{figure*}

In Fig.~\ref{fig:pdfs} we plot the unnormalized p.d.f. for the three fields, with errors (smaller than the data points) estimated from the standard deviation of the 40 simulations, and for both smoothing scales. As expected, the unmodified field is reasonably well approximated by a log-normal distribution with some additional skewness. After applying Gaussianizing transforms this distribution becomes slightly more Gaussian, with a third-order transform outperforming the second-order transform.

In the high-$k_s$ field, most of the improvement comes from the down-weighting of underdense regions (e.g. voids) and strongly overdense regions having $\delta \gtrsim 2$, and the up-weighting of mildly overdense regions having $0 \lesssim \delta \lesssim 2$. Quantitatively, we find a Gaussian fit to the p.d.f. improves after a second-order transform, reducing $\chi^2$ by 20\%. A third-order transform provides additional improvement, with $\chi^2$ reduced by 60\%.
This rough measure of Gaussianization suggest that most of the improvement arises from the subtraction of the third-order term in Equation~\eqref{eq:HM}, and a visual inspection of Fig.~\ref{fig:pdfs} supports this. The Gaussianization is extremely mild, not getting close to the one-point distribution of the linear field, whose p.d.f. (after scaling with the linear growth factor to $z=1$) we also plot in Fig.~\ref{fig:pdfs}.\footnote{Note that the fields here are permitted to have values $\delta < -1$, due to both the correction for the cloud-in-cell binning and, in the case of the linear field, the scaling with the growth factor to $z=1$.}

In the high-$k_s$ case, we find that the dimensionless skewness of the distribution reduces from $1.8$ to $1.7$ after applying a second-order transform, and to $1.5$ after a third-order transform, which should be compared with the more dramatic suppression of the large-scale folded bispectrum seen in Section~\ref{subsubsec:bispec}. The dimensionless excess kurtosis reduces from $4.1$ to $3.7$ after applying a second-order transform, and to $3.2$ after a third-order transform. Note that since these numbers are all $\gtrsim 1$ our perturbative treatment would break down if applied at the one-point level at this smoothing scale.

In the low-$k_s$ case we see much better performance of the Gaussianizing transforms, with a Gaussian fit to the second-order transformed field reducing $\chi^2$ by 80\% and the third-order transformed field by almost 100\%. The dimensionless skewness of the distribution reduces from $1.2$ to $0.5$ after applying a second-order transform, and to $0.2$ after a third-order transform, a much more dramatic effect than the high-$k_s$ case. Additionally, the dimensionless excess kurtosis reduces from $2.6$ to $0.3$ after applying a second-order transform, and to $0.2$ after a third-order transform. That the transforms appear to work much more effectively when small scales are filtered out suggests that these scales are contaminated by the poor treatment of high-$k$ modes in the raw density field, which are incorrectly handled by our method. 


The results of this section suggest that our transforms provide only mild improvement in the Gaussianity of the density field at the one-point level when including scales $k \lesssim 0.8 \, h \, \Mpc^{-1}$, but are much more effective when including scales $k \lesssim 0.2 \, h \, \Mpc^{-1}$, where the improvement is significant. It is difficult to say if this comes from a genuine enhancement of the coherence with the linear Gaussian field or an effective smoothing of the non-linear modes of the field, since we have seen that both effects are present in our transformed fields.

\subsubsection{Power spectrum variance}
\label{subsubsec:covmat}

\begin{figure}
\centering
\includegraphics[width=\columnwidth]{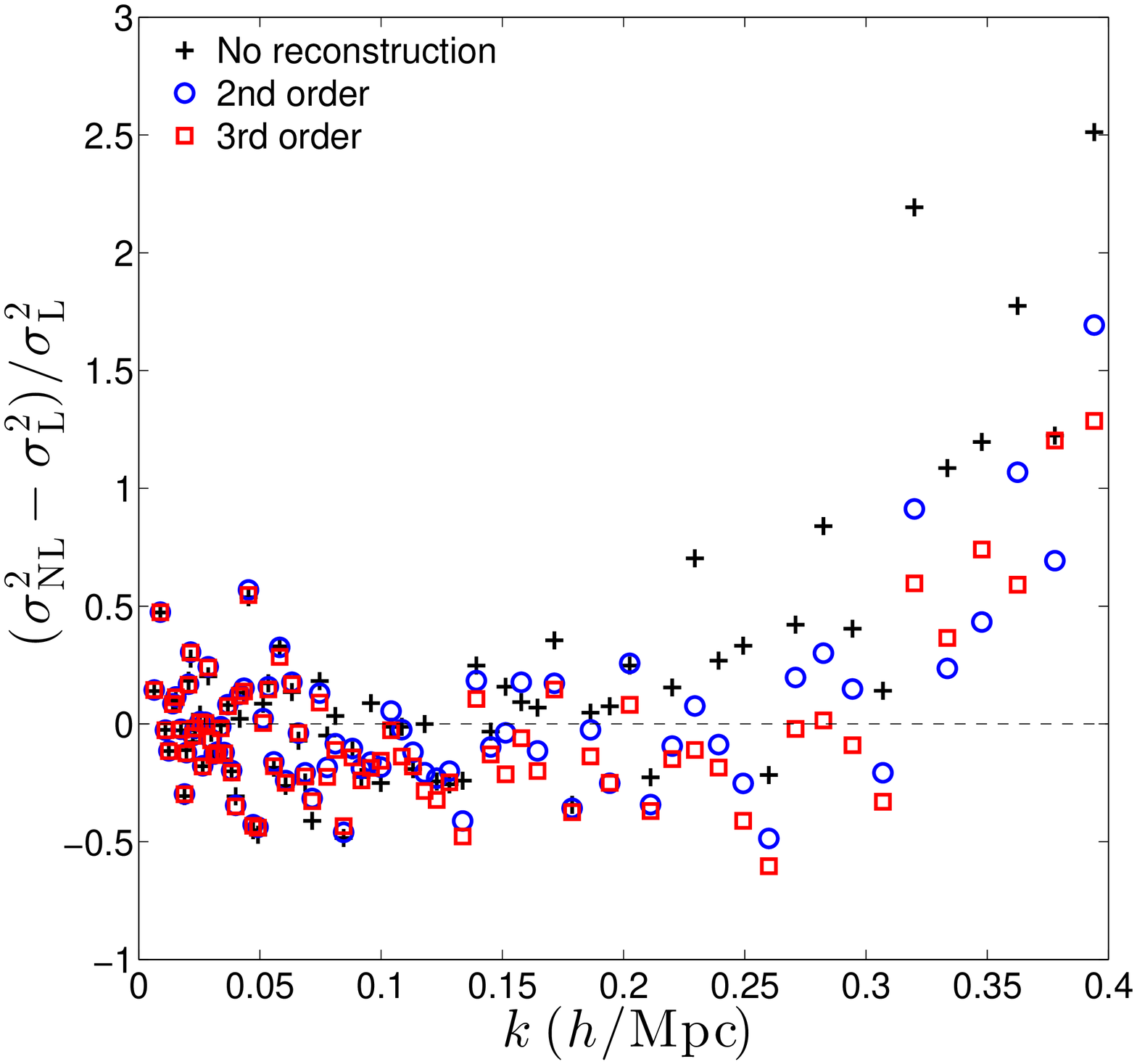}
\includegraphics[width=\columnwidth]{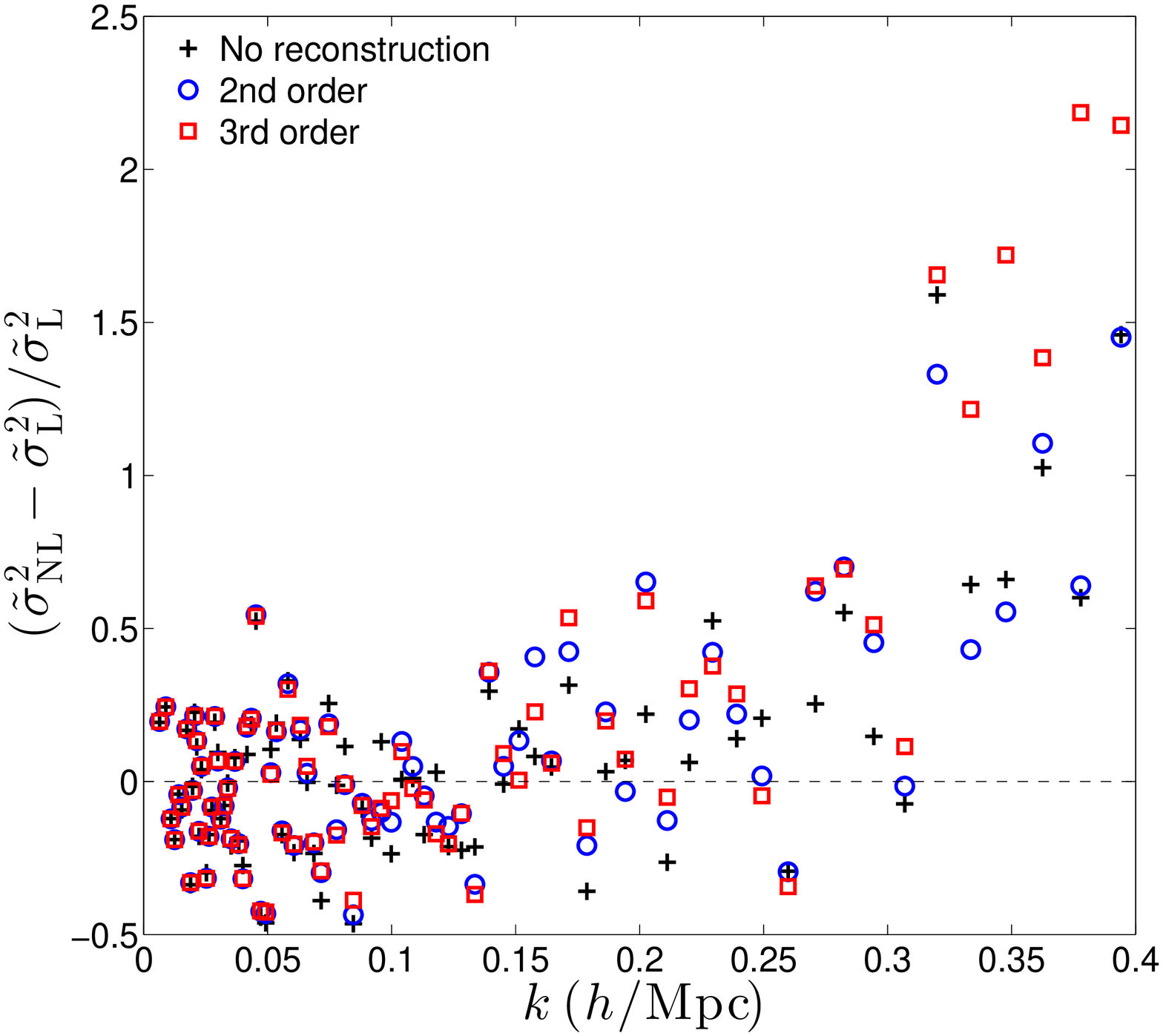}
\caption{\emph{Upper panel}: The fractional difference of the variance of the power spectrum estimates estimated from our 40 simulations at $z=1$ compared to the Gaussian prediction with the linear power spectrum, for the unmodified field (black crosses), the second-order transformed field (blue circles), and the third-order transformed field (red squares). \emph{Lower panel}: Same as top panel for the normalized variance $\tilde{\sigma}^2$.}
\label{fig:var_sims_theory}
\end{figure}

In the upper panel of Fig.~\ref{fig:var_sims_theory} we plot the fractional difference of the variance of the unmodified and transformed fields ($\sigma^2_{\rm{NL}}$) with respect to the Gaussian prediction ($\sigma^2_{\rm{L}} = 2P_L^2/N_k$, with $N_k$ the number of modes contributing to the $k$-bin). The scatter is large due to the finite number of simulations at our disposal, but some general trends can be identified. Firstly, as expected, the unmodified field displays enhanced variance on small scales due to non-linear structure formation. Secondly, there is tentative evidence that the transformed fields indeed have lower variance than the unmodified field, with a third-order transform generally outperforming a second-order transform across the non-linear scales. Note that there are no error bars on this figure, as we do not have enough simulations to accurately quantify the variance on these variance estimates. However, at leading order the points should be roughly uncorrelated between different $k$-bins, so the scatter gives an approximate measure of the errors on large scales.

How much of the decreased variance is down to an effective smoothing of the density field? To answer this, in the lower panel of Fig.~\ref{fig:var_sims_theory} we plot the fractional difference of the normalized variances defined by $\tilde{\sigma}^2 = \sigma^2/P^2$, which (for a diagonal covariance matrix) is proportional to the inverse signal-to-noise-squared on the power spectrum in each $k$-bin. The small reduction in variance is now below the scatter in the measurements, with a potential \emph{increase} in the variance of the third-order transformed field on the smallest scales. This tells us that any reduction in variance comes with a reduction in the power spectrum, such that the signal-to-noise is roughly unchanged. Any increase in signal-to-noise brought by our transforms is therefore at least as small as the (large) error bars on these variance measurements.

%
%
Ideally the transforms should also reduce the off-diagonal terms of the power spectrum covariance matrix, since these are sourced by the trispectrum of the field. We computed these terms, but the number of simulations at our disposal proved insufficient to measure any changes with high enough statistical significance.

The results of this section suggest that our transforms do indeed reduce the variance of power spectrum estimates, although at a level similar to the suppression of the power spectrum, suggesting that both signal and noise have effectively been reduced by transforms. We note however that the enhancement of the correlation with the initial field suggests the effective smoothing is accompanied by an increase in information content.

\subsubsection{Baryon Acoustic Oscillations}
\label{subsubsec:bao}

In order to make more quantitative statements from our limited number of simulations, in this section we focus on linearizing the wiggles in the real-space dark matter power spectrum imparted by baryon acoustic oscillations (BAO). Our perturbative approach should be of great use here as BAO scales are in the linear or quasi-linear regime, and the effects of non-linear structure formation are mild and mostly of linear origin~\citep{2006PhRvD..73f3519C,2007ApJ...664..660E,2008PhRvD..77b3533C,2012PhRvD..85j3523S,2014JCAP...02..042S}.

We apply the Gaussianization transform Equation~\eqref{eq:HM} to three pairs of simulated density fields. Each member of a pair has the same initial random seed but different initial power spectra, one having the fiducial linear $P_L(k)$ and the other having the same broadband shape but with its BAO wiggles removed with a low-pass filter\footnote{To create the wiggle-free spectrum we take the Fourier Transform of $P_L(k)$ in log-space and apply a low-pass filter to remove the BAO before returning to real space. We then create the smooth spectrum by stitching in the original spectrum above and below the BAO scale in order to removed edge effects induced by the Fourier Transform.}. Relative differences between `wiggle' and `no-wiggle' power spectra are therefore free of cosmic variance at leading order, greatly reducing the number of simulations required to draw reliable quantitative conclusions about our method. We confirmed that all our results were stable to increasing the number of paired density fields from three to four. The input parameters of the simulations were the same as in Section~\ref{subsubsec:band}, and we again study real-space density fields at $z=1$ in this section.

\begin{figure}
\centering
\includegraphics[width=\columnwidth]{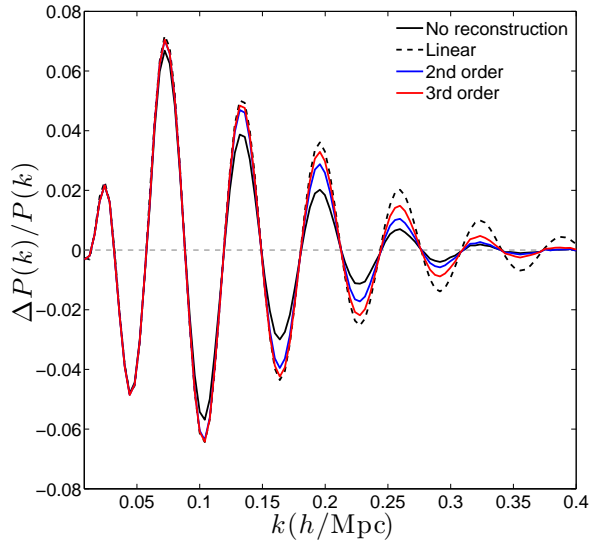}
\caption{Fractional difference of the wiggle and no-wiggle power spectrum in real space at $z=1$ for the unreconstructed field (black solid, lowest amplitude curve), linear field (black dashed), the second-order reconstructed field (blue solid, middle amplitude curve) and the third-order reconstructed field (red solid, highest amplitude curve).}
\label{fig:dmbaopdiff}
\end{figure}

In Fig.~\ref{fig:dmbaopdiff} we plot the fractional difference of the wiggle and no-wiggle matter power spectrum for our second-order and third-order Gaussianization schemes given by Equation~\eqref{eq:HM}. This figure clearly demonstrates that our method reconstructs the linear BAOs on scales $k \lesssim 0.3 \, h \, \Mpc^{-1}$, undoing a large fraction of the damping of the wiggles caused by non-linearity. The frequency and phase of the wiggles is preserved by the transformation on these scales, showing that our transforms do not bias the signal, unlike the case of the broadband power\footnote{This also implies that the peak of the correlation function is not significantly shifted by our transformation, which we confirmed qualitatively with a simple estimate for the correlation function formed by Fourier transforming the power spectrum estimates.}. Furthermore, the third-order transformation outperforms the second-order transformation, suggesting that our perturbative expansions are well-behaved in this regime.

To quantify this sharpening of the peaks, we compute the cumulative squared signal-to-noise on the wiggles below a given scale, $(S/N)^2(<k)$, defined as
\begin{equation}
(S/N)^2(<k) \equiv \sum_{k_b < k} \frac{N(k_b)(P_w(k_b) - P_{nw}(k_b))^2}{2P_{nw}(k_b)^2},
\label{eq:sn}
\end{equation}
where $N(k_b)$ is the number of modes in a bin centred at $k_b$, and the sum is over bins. This definition assumes Gaussian noise on the power spectrum estimates, which should be accurate on these scales~\citep{2009ApJ...700..479T,2015PhRvD..92l3522S}\footnote{Even if this were not true in detail, the variance on the transformed power spectrum should be more Gaussian than the raw power spectrum estimates.}. Note that the paired simulations with common initial conditions dramatically reduces the noise in this statistic due to the ratio in Equation~\eqref{eq:sn}, and our results change negligibly if we use two pairs instead of three.

\begin{figure}
\centering
\includegraphics[width=\columnwidth]{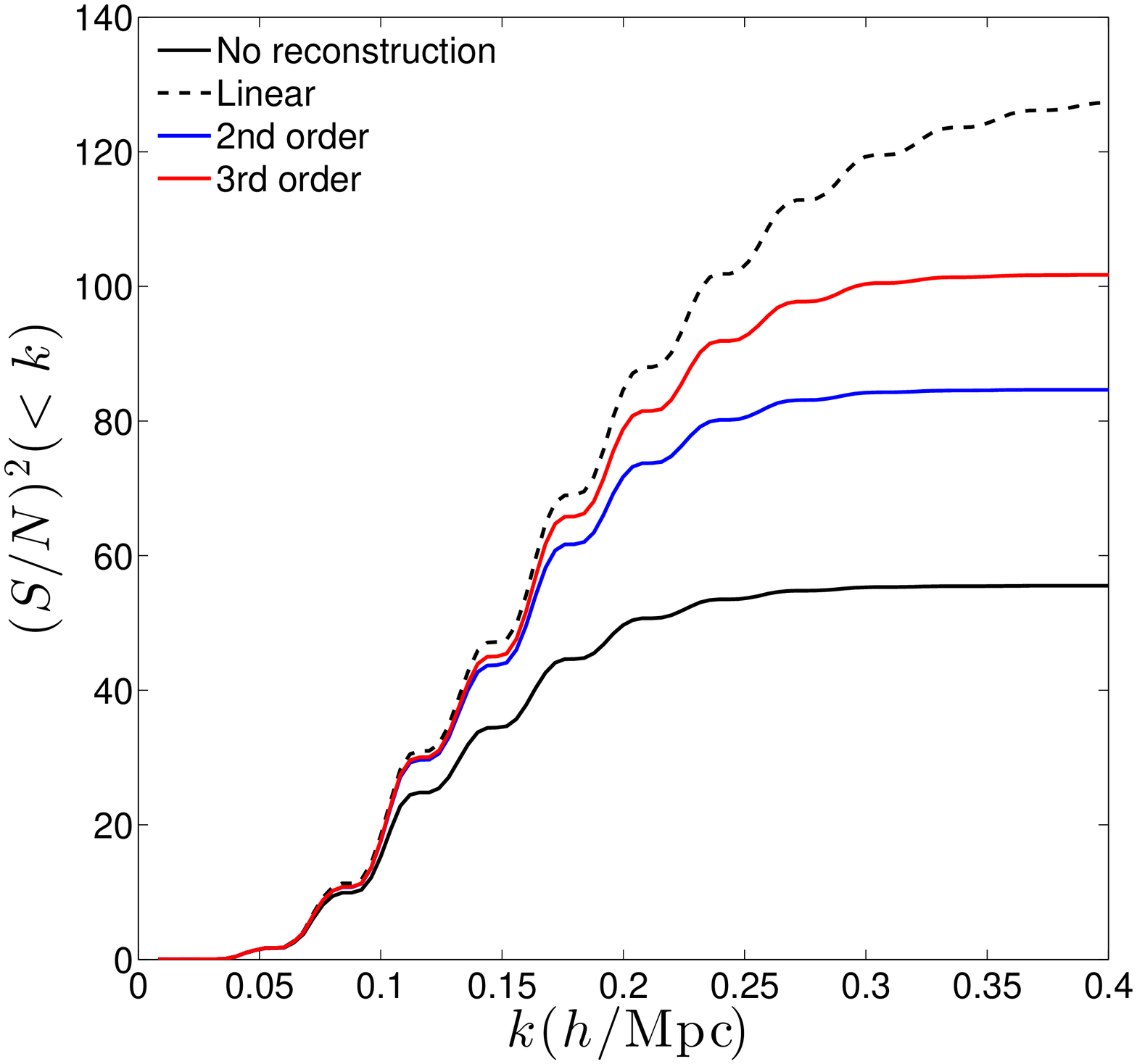}
\caption{Cumulative square signal-to-noise in real space at $z=1$ for the unreconstructed field (black solid, lower curve), linear field (black dashed), the second-order reconstructed field (blue solid, middle curve) and the third-order reconstructed field (red solid, upper curve).} 
\label{fig:dmbaosn}
\end{figure}

In Fig.~\ref{fig:dmbaosn} we plot the quantity $(S/N)^2(<k)$. The sharpening of the peaks is now quantified as increase in the $S/N$ as a second-order and then a third-order reconstruction are applied. Most of the improvement occurs on scales $k \lesssim 0.25 h\, \mathrm{Mpc}^{-1}$, since on smaller scales the significant non-linear growth washes out the BAO wiggles to the extent that our perturbative transforms cannot recover linear information. The total increase in $S/N$ (i.e. the square-root of the asymptotic values in Fig.~\ref{fig:dmbaosn}) is roughly 20\% for the second-order transform and 35\% for the third-order transform. In comparison, the Gaussian linear field has over 50\% more information on the wiggles than the raw field, demonstrating that our linearizing transform captures a significant proportion of the total information available. 

We have thus seen that our Gaussianizing transform successfully recovers a large fraction of the linear real-space BAO signal washed out by non-linear structure formation. The method produces broadband power spectrum estimates which are biased with respect to the linear field, but we have seen evidence that the variance of these estimates is reduced. We have restricted to $z=1$ where our perturbative approach is expected to be reasonably accurate on BAO scales, but note that worse or improved performance could be expected at lower or higher redshifts respectively.


\section{Application to 21\,cm intensity mapping}
\label{sec:21cm}

We have seen that the Gaussianizing transform of Equation~\eqref{eq:HM} successfully removes a large fraction of the non-linear smoothing of the BAO wiggles in the real-space matter power spectrum. In this section, we test our simple method further by applying it to mock realizations of a 21\,cm intensity map. Several 21\,cm BAO reconstruction methods based on perturbative models for the 21\,cm brightness temperature have recently been tested in the literature~\citep{2016MNRAS.456.3142S, 2016MNRAS.457.2068C, 2017JCAP...09..012O}. Since we have made the approximation that the bispectrum and trispectrum of the dark matter density are given by their tree-level expressions, our method is not expected to be as effective at reconstruction as those based on Lagrangian perturbation theory such as the pixel remapping approach of~\citet{2017JCAP...09..012O}. This is because the main physical effects of BAO smoothing are captured by leading order Lagrangian perturbation theory (i.e. the Zeldovich approximation) in a way which cannot be reproduced with a finite number of terms in the Eulerian framework in which we work, due to the non-perturbative mapping between Lagrangian and Eulerian space. Nevertheless, it is useful to compare the performance of Lagrangian methods applied to 21\,cm pixels with an Eulerian approach to test this reasoning quantitatively.

21\,cm intensity maps have several complicating features over the real-space dark matter density field considered in Section~\ref{sec:dm}. Firstly, since the field is now a continuous map of brightness temperature fluctuations rather than discrete number counts, we have to account for the loss of angular resolution due to the finite size of the interferometer or receiving dish (see, e.g.~\citealt{2015ApJ...803...21B} for a discussion of 21\,cm systematics). On the other hand, due to the clean spectral signature of the 21\,cm line, intensity maps typically have excellent radial resolution allowing for accurate redshifts to be obtained. The caveat to this is a loss of large-scale radial information due to the large foregrounds which need to be subtracted. Foregrounds typically have structure on frequency scales corresponding to small values of $k_{\parallel}$, and their removal leads to the loss of these modes. Secondly 21\,cm intensity maps contain noise from the detector, which renders the beam deconvolution non-trivial. Thirdly, as with galaxy redshift surveys, observations of intensity maps occur in redshift space rather than real space. Finally, since neutral hydrogen on cosmological scales is expected to reside within dark matter haloes, the distribution 21\,cm brightness temperature fluctuations is biased with respect to the underlying dark matter.

\begin{figure*}
\centering
\includegraphics[width=0.8\columnwidth]{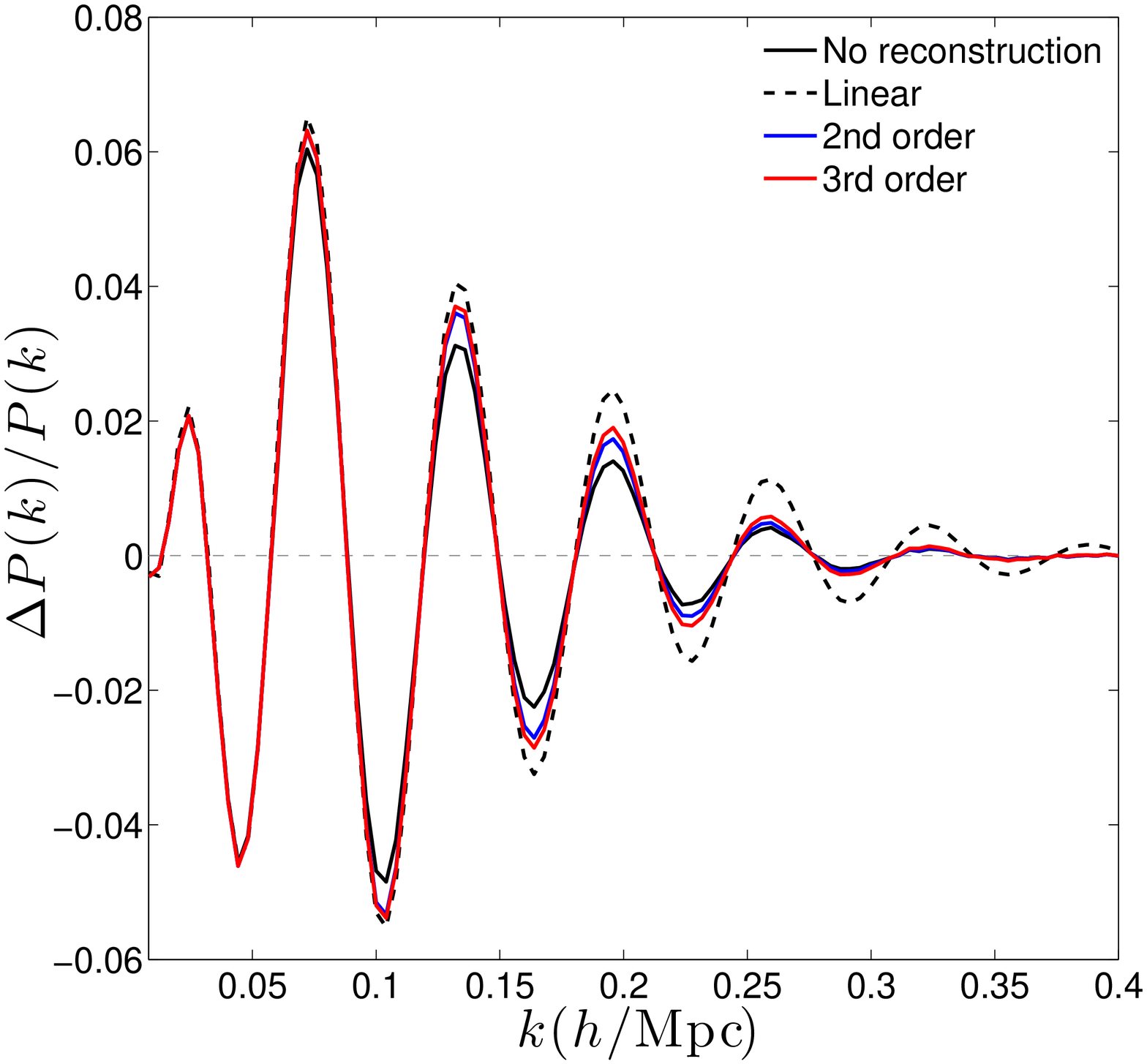}
\includegraphics[width=0.8\columnwidth]{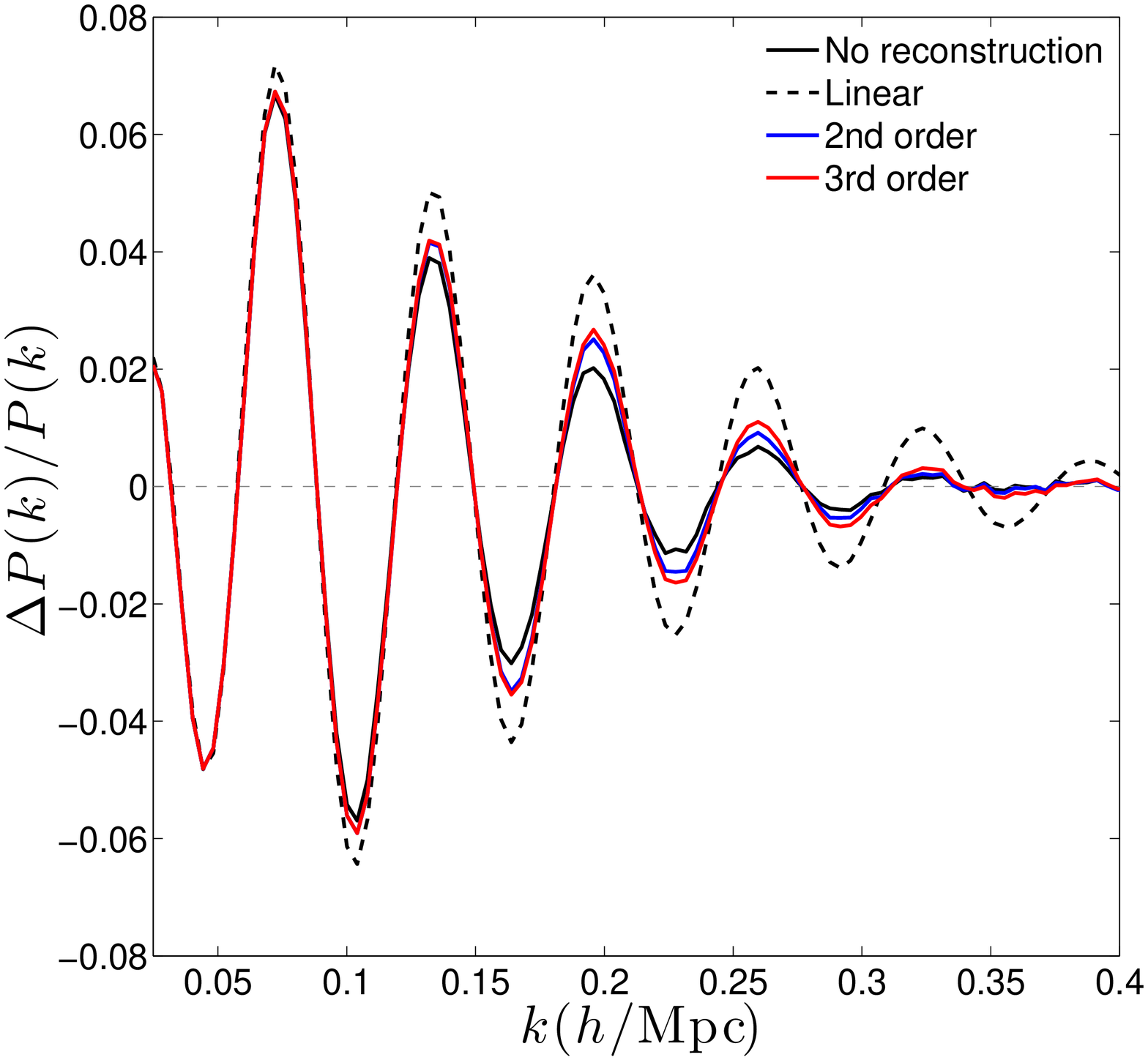}
\includegraphics[width=0.8\columnwidth]{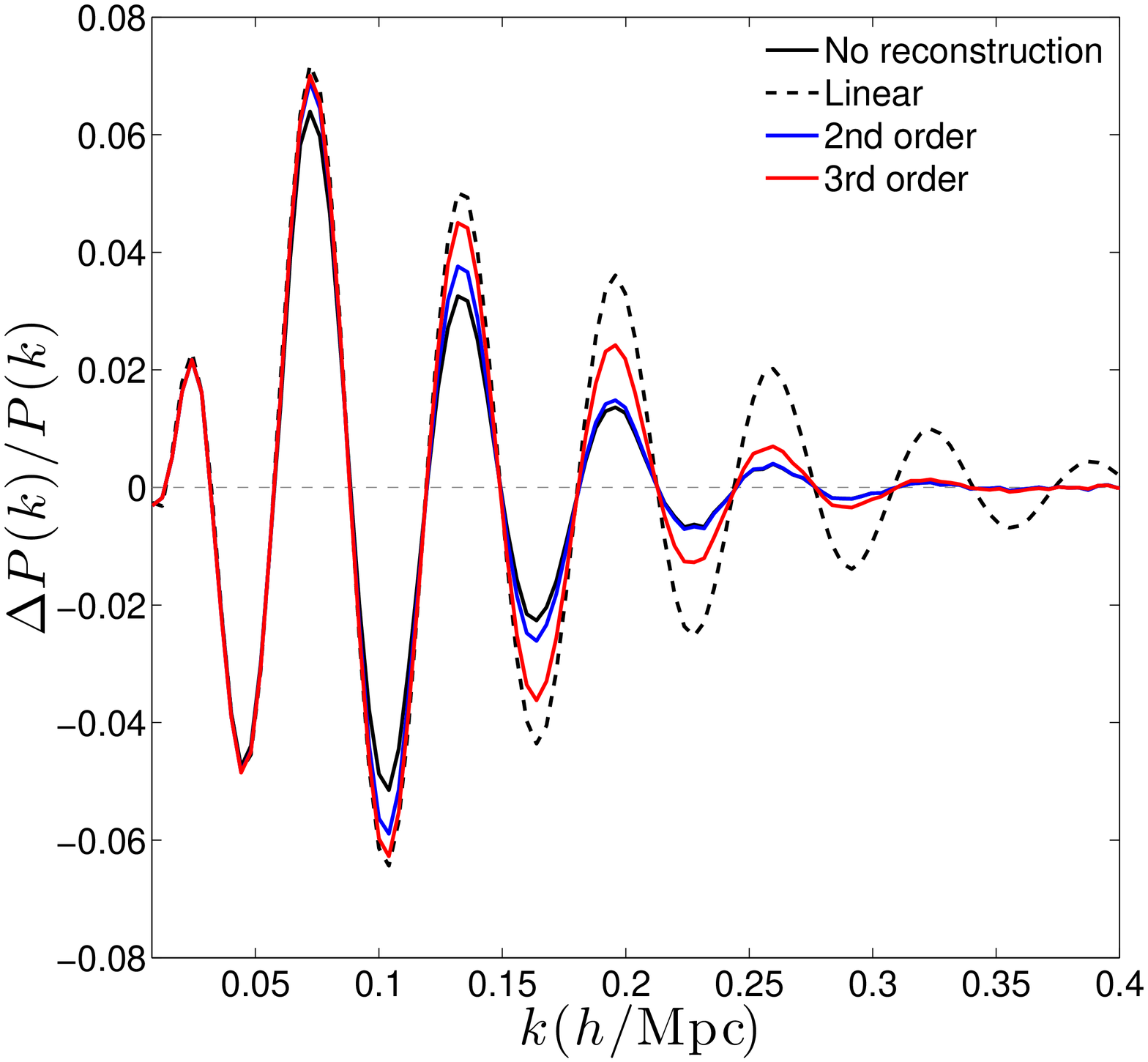}
\includegraphics[width=0.8\columnwidth]{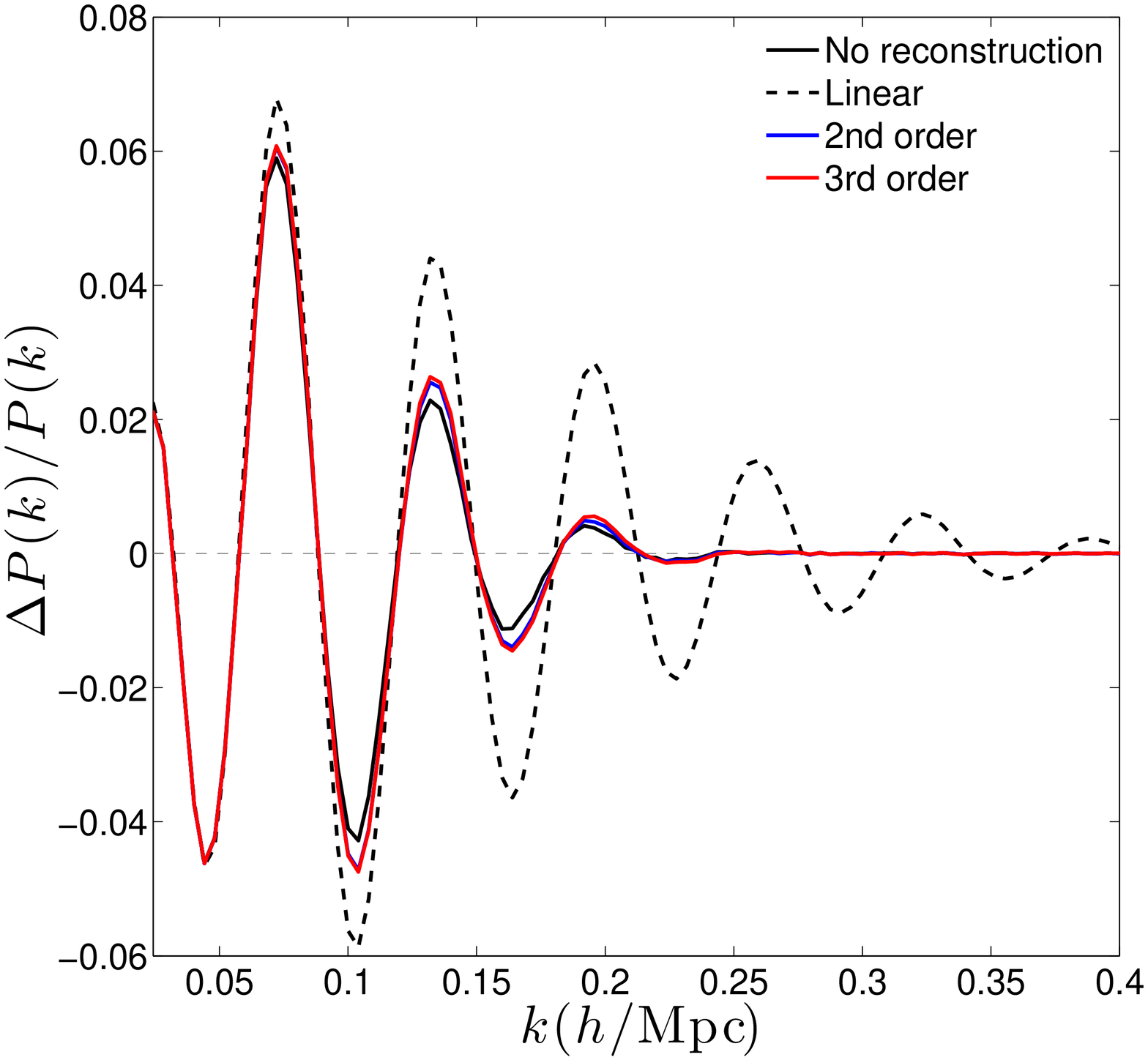}
\caption{\emph{Top left}: Fractional difference between the wiggle and no-wiggle power spectra at $z=1$ in real-space with pixel noise for the unreconstructed field (black solid, lowest amplitude curve), linear field (black dashed), second-order reconstructed field (blue solid, middle amplitude curve) and third-order reconstructed field (red solid, highest amplitude curve). \emph{Top right}: Same, for the real-space field with $k$-space filters and smoothing applied (note the change in the range of the horizontal axis due to the high-pass filter). \emph{Bottom left}: Same, for the redshift-space field. \emph{Bottom right}: Same, when all three effects are implemented.}
\label{fig:21cmpdiff}
\end{figure*}

To account for these complications, we build very simple mock intensity maps from our pairs of simulated dark matter density fields at $z=1$. We consider three separate scenarios: a redshift-space density field, a real-space density field with beam-smoothing, finite radial resolution and a cut on $k_{\parallel}$ to account for foregrounds, and a real-space density field with uncorrelated statistically homogeneous Gaussian noise added to each pixel. Finally we consider a scenario in which all three of these effects are included (with the noise added after beam smoothing, and the beam smoothing added after the shift to redshift space). For simplicity we do not include the effects of bias or more complicated sampling of Fourier space due to scanning strategy or foreground `wedge' effects~\citep{2016MNRAS.456.3142S}. Our maps are thus highly idealized, but allow for the evaluation of the major complicating factors to be assessed individually. We choose noise and smoothing specifications to be similar to those expected from near-term intensity mapping experiments such as CHIME~\citep{2014SPIE.9145E..4VN}. Beam smoothing in the angular direction is implemented with a Gaussian in the $k_{\perp}$ direction having size $10 \, \mathrm{Mpc} \, h^{-1}$, and the loss of radial information due to finite frequency resolution is achieved by multiplying the Fourier-space density field with a Gaussian in $k_{\parallel}$ of size $3\, \mathrm{Mpc} \, h^{-1}$. We also apply a hard high-pass filter removing modes having $k_{\parallel} < 0.02 \, h \, \mathrm{Mpc}^{-1}$ to account for foregrounds. When adding noise to each Fourier space pixel we generate a random Gaussian realization with power spectrum $P_N = 330 \, \mathrm{Mpc}^3 \, h^{-3}$, corresponding to an effective source number density of $\bar{n} = 3 \times 10^{-3} \, h^{-3} \, \mathrm{Mpc}^{-3}$, consistent with that used in~\citet{2016MNRAS.457.2068C}. To boost the statistical power of our BAO forecasts, we use the same noise realization within each pair of wiggle/no-wiggle simulations. Note that noise does not bias the signal since it contributes equally to the wiggle and no-wiggle density fields, but it does contribute to the variance.

\begin{figure*}
\centering
\includegraphics[width=0.8\columnwidth]{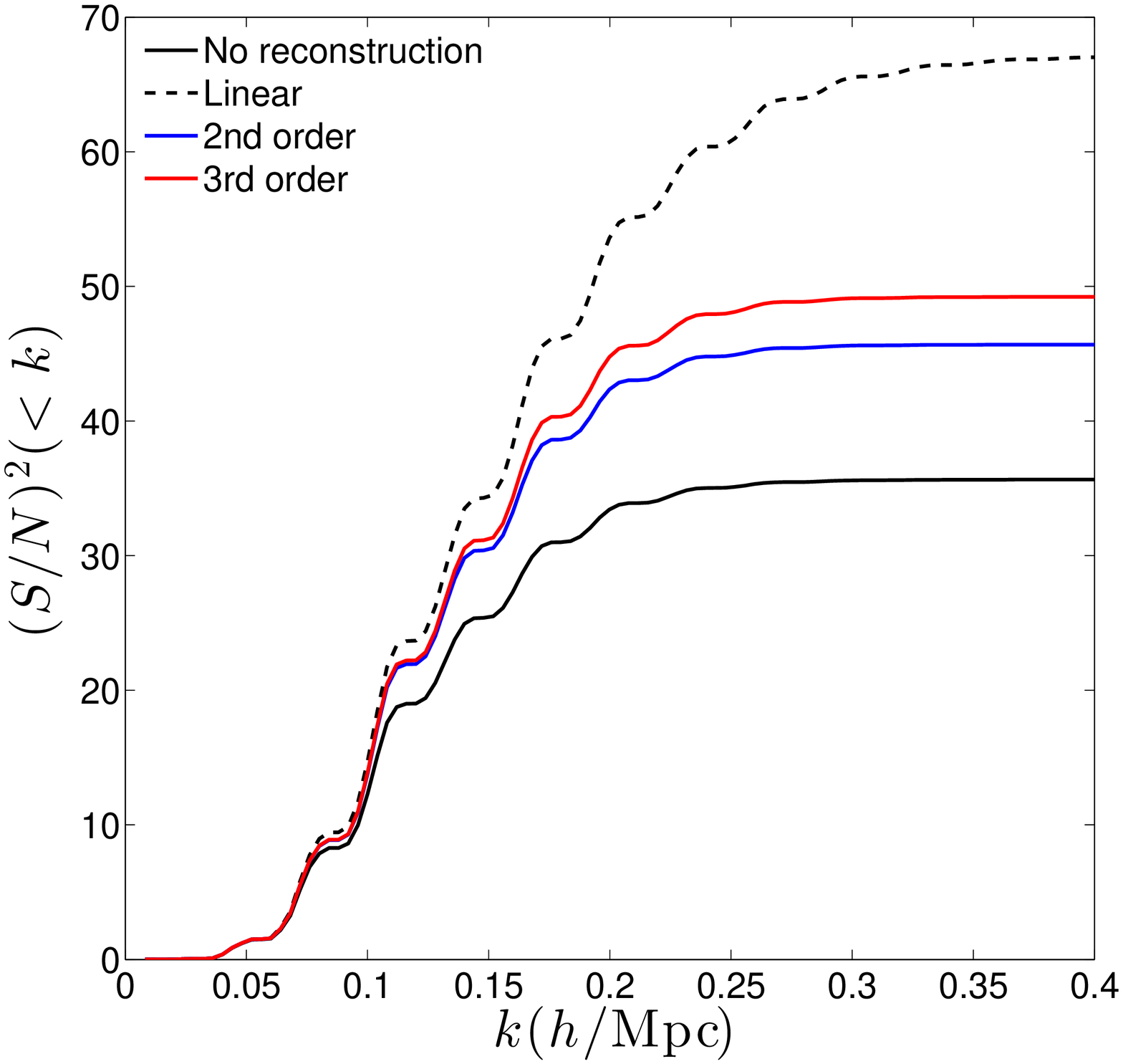}
\includegraphics[width=0.8\columnwidth]{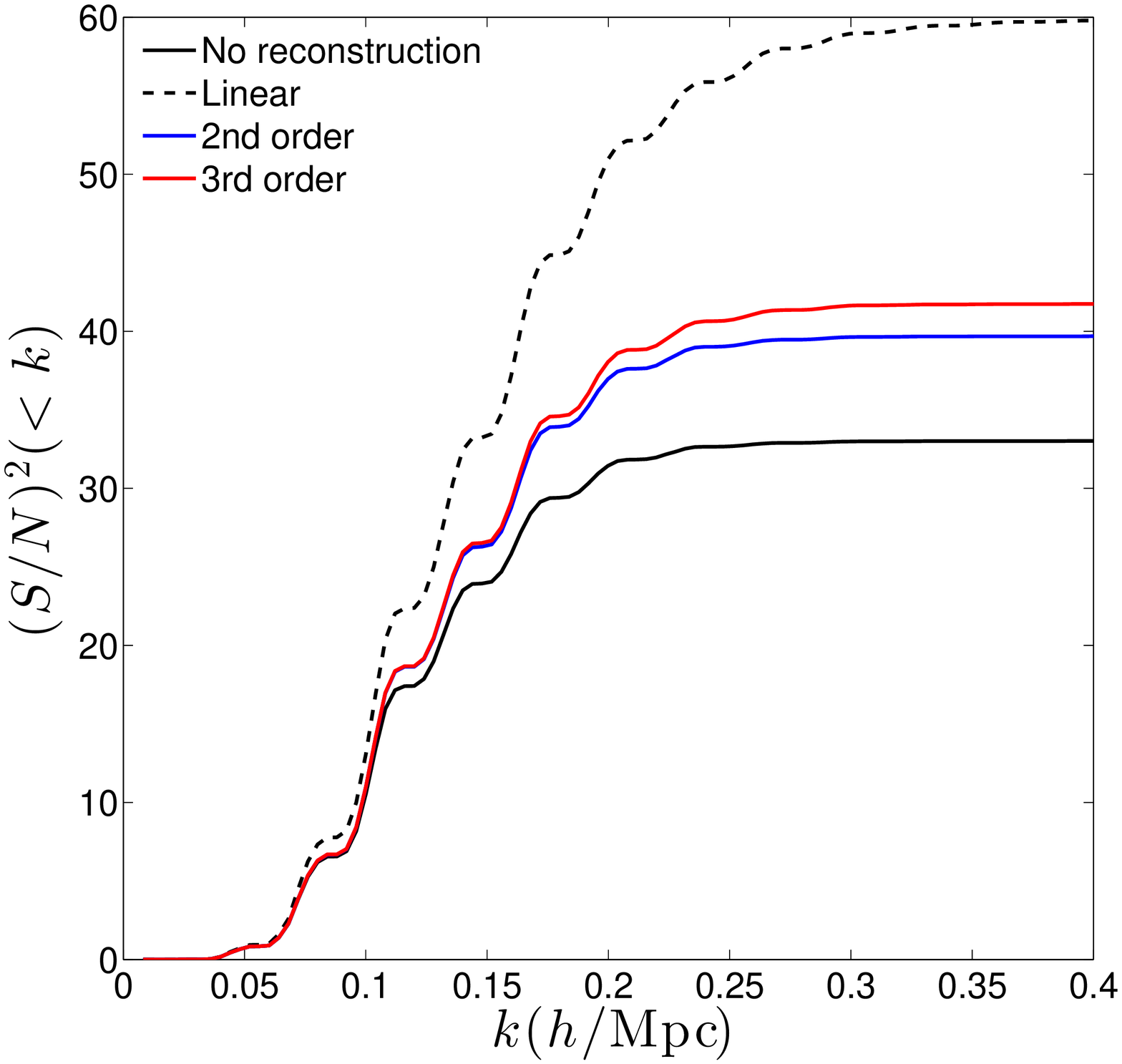}
\includegraphics[width=0.8\columnwidth]{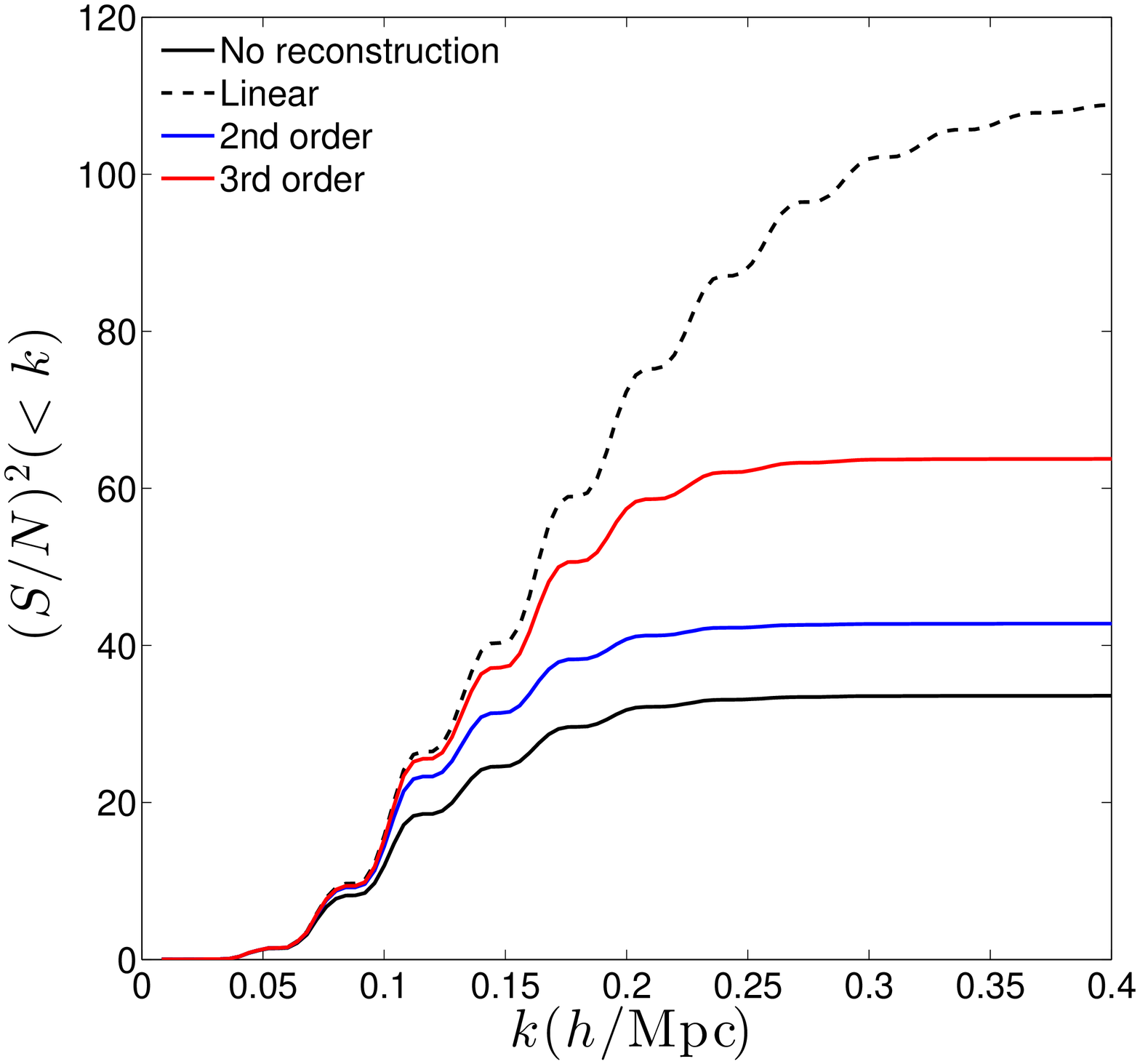}
\includegraphics[width=0.8\columnwidth]{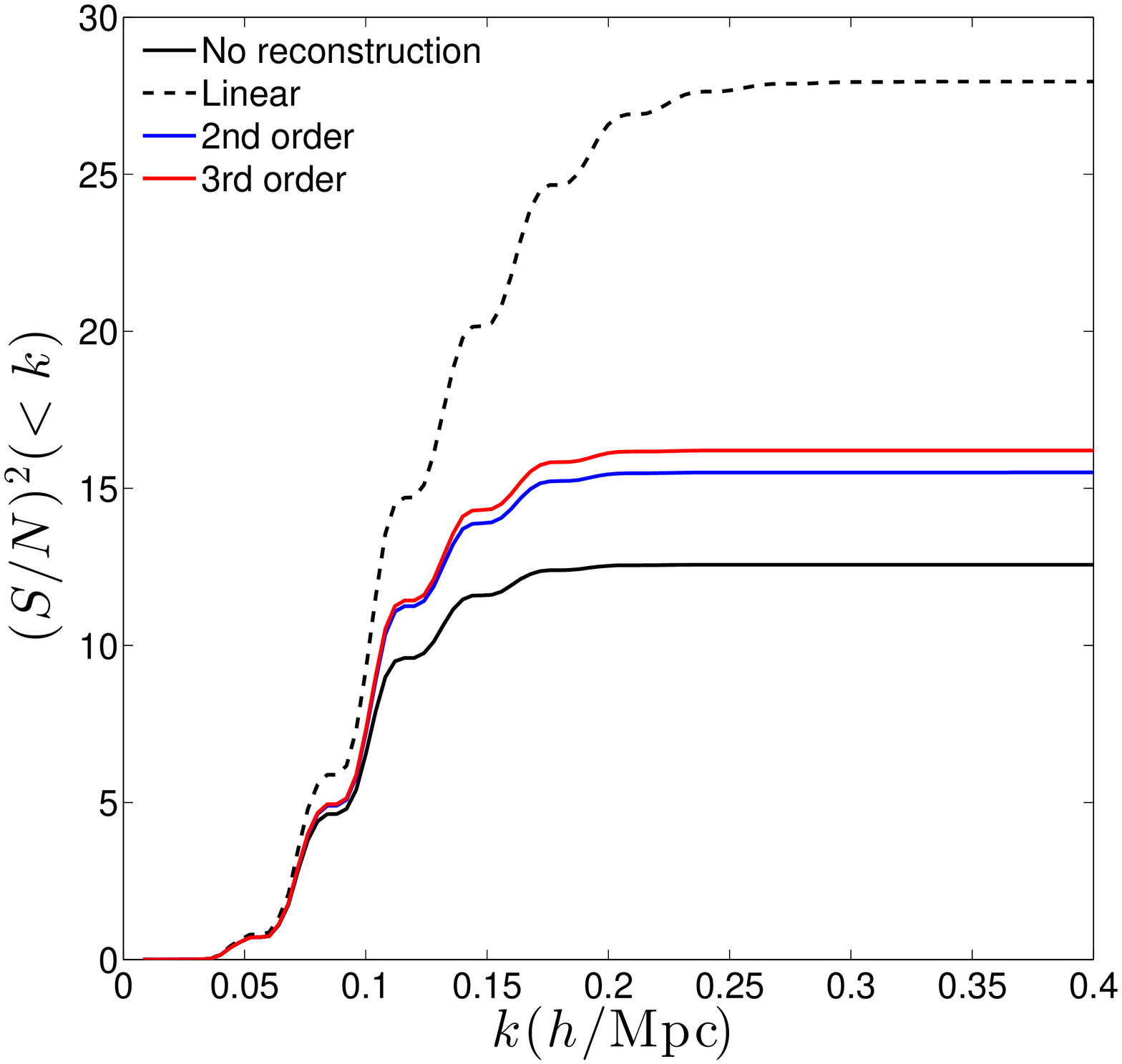}
\caption{\emph{Top left}: Cumulative squared signal-to-noise of the BAO wiggles at $z=1$ in real-space with pixel noise for the unreconstructed field (black solid, lower curve), linear field (black dashed), second-order reconstructed field (blue solid, middle curve) and third-order reconstructed field (red solid, upper curve). \emph{Top right}: Same, for the real-space field with $k$-space filters and smoothing applied (note the change in the range of the horizontal axis due to the high-pass filter). \emph{Bottom left}: Same, for the redshift-space field. \emph{Bottom right}: Same, when all three effects are implemented.}
\label{fig:21cmsn}
\end{figure*}

Since some of our maps now contain noise, the Gaussianizing transform Equation~\eqref{eq:wrongskew} must be modified to account for the fact that the power spectrum of the observable has changed. Using the formalism of Section~\ref{sec:gauss}, it is straightforward to show that all our expressions may be generalized to include noise by modifying the power spectrum as $P_L(k) \rightarrow P_L(k) + P_N$, where we recall that $P_L(k)$ is the linear power spectrum. This is equivalent to multiplying the density field by the Wiener filter $P_L(k)/[P_L(k) + P_N]$ prior to constructing the non-linear combination required to remove the leading-order non-Gaussianity. In the presence of noise, redshift-space distortions (RSDs) and beam smoothing, the appropriate filter becomes $P_L(k)/[P_L(k) + (1+f\mu^2)^2B(k)P_N]$, where $f$ is an estimate of the growth factor, $\mu$ is the cosine of the angle between the wavevector and the line of sight~\citep{1987MNRAS.227....1K} and $B(k)$ is the beam smoothing function in Fourier space.

In Fig.~\ref{fig:21cmpdiff} we plot the fractional differences in the monopole power spectra of the wiggle and no-wiggles intensity maps at $z=1$ for each of the four scenarios detailed above. Reassuringly we see that in all cases a second-order transform outperforms the unreconstructed field, with a third-order transform performing even better. The presence of pixel noise boosts the broadband power and reduces the contrast of the BAO wiggles, reducing the amplitude of the curves in the top-left panel of Fig.~\ref{fig:21cmpdiff}. Noise is most destructive on small spatial scales due to its white spectral shape, and washes out linear modes which the reconstruction methods need to capture linear power on larger scales. The result of this is that a third-order reconstruction improves little over a second-order reconstruction. The loss of small scales is more detrimental in the case of beam smoothing (top-right panel of Fig.~\ref{fig:21cmpdiff}), with the poor angular resolution of our fiducial 21\,cm experiment limiting the power of both reconstruction methods to linearize the intensity maps. Our tree-level expressions for the bispectrum and trispectrum do not include RSDs, and instead we treat them as contaminants and assess their impact on the reconstructions. When the fields are shifted to redshift space, the bottom-left panel of Fig.~\ref{fig:21cmpdiff} shows that a third-order reconstruction outperforms a second-order reconstruction, which barely improves over the case of no reconstruction. When all three effects are switched on, we find that our linearization routines still offer improvements over the raw intensity map, but the effects are very small (bottom-right panel of Fig.~\ref{fig:21cmpdiff}). This suggests that our method requires more detailed modelling of redshift-space and beam effects. The difficulty of modelling non-linear RSDs in Eulerian space exemplifies one of disadvantages of Eulerian BAO reconstruction methods over Lagrangian techniques.


In Fig.~\ref{fig:21cmsn} we plot the cumulative squared signal-to-noise for the BAO wiggles from the intensity maps, along with the linear (Gaussian) result. The modest improvement of the third-order reconstruction over the second-order reconstruction is quantified as a small improvement over the total signal-to-noise in the BAO wiggle, with the greatest improvement being for the redshift-space field in the absence of noise and beam smoothing. The top-right panel of Fig.~\ref{fig:21cmsn} demonstrates that the loss of large-scale radial modes due to foreground subtraction does not have a large impact on the total signal-to-noise of the BAO wiggles, due to the mismatch of the relevant scales. When we include noise, RSDs, and beam smoothing, we find similar total signal-to-noise from both our reconstruction methods, with a roughly 15\% increase over the unreconstructed field compared to the 50\% increase potentially available based on a Gaussian prediction.




\section{Conclusions}
\label{sec:conc}

In this work we have derived conditions which non-linear and non-local transformations of cosmological fields must satisfy in order to possess Gaussian statistics. The fundamental assumption behind our formalism is that the field is weakly non-Gaussian, in the sense that its cumulants or polyspectra become successively smaller at each order. This assumption is valid for most cosmological observables on sufficiently large scales, where the non-Gaussianity induced by non-linear structure formation has not yet become strong, and is also valid at sufficiently high redshifts. The scales at which our perturbative expansion breaks down is uncertain however, and requires numerical simulations to be evaluated.

The constraints we derived suggest a Gaussianizing transform with a particularly simple form, given by Equation~\eqref{eq:wrongskew}. Unlike other linearizing or Gaussianizing transforms in the literature, this expression is both non-local and written entirely in terms of the polyspectra of the field. Thus, we do not require a detailed perturbative model for the observable, and these polyspectra could in principle be estimated from simulations. Our formalism is therefore expected to be particularly useful for maps of the weak lensing shear.

Although a perturbative model for the field is not required, quick and straightforward tests of our formalism can be made when a choice of transform based on a further perturbative expansion of the cumulants is made, Equation~\eqref{eq:HM}. We found that in this case our formalism effectively subtracts off the leading non-linear parts of the observable. With N-body simulations we showed that the bispectrum of the dark matter density field is reduced.  We have also shown that the one-point probability distribution of the density field at $z=1$ is closer to a Gaussian after applying the transforms, and that the correlation with the initial Gaussian field is enhanced, suggesting that information gain is achieved by transforming the field to a form more coherent with its initial linear state. We also saw that the power spectrum mean and variance are suppressed by the transforms such that the signal-to-noise is roughly unchanged, although we did not have enough simulations to quantify this rigourously.

The poor performance of our method in the broadband power spectrum is somewhat surprising given the enhanced Gaussianity and correlation with the initial field seen in the previous sections. One possible explanation of these discrepancies is that the various smoothing operations in our transform and the mode-coupling it induces in the non-linear field have conspired to produce a field of the approximate form $W(k)\delta_L(\vk)$ instead of $\delta_L(\vk)$, where $W(k)$ is a smooth function approximately independent of $\delta_L(\vk)$. Such a field would display Gaussianized and linearized properties passing the test of the previous sections, but would not have a linearized power spectrum. A perturbative expansion of Equation~\eqref{eq:HM} suggests that $W^2(k)$ is given by one-loop integrals over the linear power, the smoothing kernel $S(k)$, and the $F_2$ perturbation theory kernel. This suggests that a more optimal choice for $S(k)$ could be found to minimize the difference between the transformed power and the linear power. Indeed, choosing a logistic function for $S(k)$ with a sharp transition does improve the linearity of the power spectrum estimates, but at the cost of increasing the variance. We defer a more detailed investigation of this issue to a future work.

We also saw that our Gaussianizing transform does a good job of reconstructing linear BAOs which are damped by non-linear structure formation. We found that our third-order transformation increases the total signal-to-noise of the BAO wiggles by 35\%, compared to the total available information which is 50\% greater than the unreconstructed field in real-space at $z=1$.

Since our transform works at the level of pixels rather than galaxies or discrete tracers, it is interesting to investigate observable for which the positions of individual tracers are not available, such as a 21\,cm intensity map. We created toy realizations of intensity maps by shifting the particles in our dark-matter-only simulations to redshift space, creating density fields, smoothing with beams to mimic the finite angular and frequency resolution of a near-term experiment, removing large-scale radial modes to account for foreground subtraction, and finally adding Gaussian pixel noise. We found that this final stage required the maps to be Wiener filtered prior to being Gaussianized, and saw that linear BAO information can still be obtained even when all these complicating effects were switched on. However, improvement was very modest (15\% increase in total signal-to-noise), suggesting that more detailed modelling of redshift-space distortions is required for this method to improve. We also restricted tests to $z=1$, noting that non-linearity in the maps is greater at lower redshifts, although an experiment such as CHIME cannot observe at wavelengths corresponding to redshifts below $z=0.8$, so our choice of redshift is justified. The formalism of this work could equally well be applied to low-redshift surveys.

Despite the simplifying assumptions detailed above, our results are encouraging and suggest that the transform of Equation~\eqref{eq:wrongskew} be tested in more detail. In particular, it would be interesting to investigate its performance on mock weak lensing shear maps for comparison with more ad-hoc local transformation such as~\citet{2011MNRAS.418..145J}. Complicating factors such as masking and inhomogeneous noise would also have to be incorporated, although these could be handled by applying the relevant mixing matrices to the Fourier-space fields.

One interesting consequence of our Gaussianizing transform is that it provides a route to modelling the fully non-Gaussian multivariate likelihood function. This quantity is usually modelled as a Gaussian (justified on large scales) or log-normal~\citep{1991MNRAS.248....1C, 1994ApJ...435..536C, 2001ApJ...561...22K, 2017MNRAS.466.1444C}. However, a principled and physically-motivated expression for the likelihood has so far proved elusive (see~\citealt{2015MNRAS.449L.105B, 2016MNRAS.460.1529U, 2017arXiv170606645S} for recent progress). Modelling this distribution is of high importance, since measuring it from simulations is extremely challenging due to the high dimensionality of the problem. It is also not known how accurate the likelihood function needs to be for future Stage-IV surveys to obtain precision constraints on dark energy - incorrectly imposing Gaussianity on the likelihood could bias parameter inferences. Furthermore, a Bayesian approach to inferring constraints on cosmological parameters requires a functional form for the likelihood to be specified (see, e.g.~\citealt{2016MNRAS.455.4452A}), which motivates more accurate modelling including bispectrum and trispectrum terms. 

By transforming a Gaussian distribution using the transform of Equation~\eqref{eq:wrongskew}, we can form a non-Gaussian, multivariate distribution which is everywhere finite and positive-definite, and which should converge to the true likelihood on sufficiently large scales. Such a distribution would model the probability of the density field given the power spectrum, bispectrum and trispectrum, and could thus be used to form a joint posterior on these polyspectra which could be sampled from. In the one-dimensional case and returning to the notation of Section~\ref{subsec:gengauss}, this distribution would be given by $P_X(X) = P_Y[Y(X)]\lvert \mathrm{d}Y/\mathrm{d}X \rvert$, and taking $P_Y$ as Gaussian and using Equation~\eqref{eq:1Dgtrans} we find
\begin{align}
&P_X(X) \approx \biggl \lvert 1 - \frac{\kappa_3}{3}X - \frac{\kappa_4}{8}(X^2-1) + \frac{\kappa_3^2}{36}(12X^2-7) \biggr \rvert \nonumber \\
&\times \exp \left\{-\frac{1}{2} \left[X - \frac{\kappa_3}{6}(X^2-1) \vphantom{\frac{(a)^{b^2}}{b}} \right. \right. \nonumber \\
& \left. \left. - \frac{\kappa_4}{24}(X^3-3X) + \frac{\kappa_3^2}{36}(4X^3-7X)\right]^2 \right \}.
\label{eq:like}
\end{align}
The multivariate generalization of Equation~\eqref{eq:like} should provide a likelihood suitable for simultaneously inferring power spectra and maps in a Bayesian hierarchical framework. We defer detailed investigation of this distribution to a future work.

\section{Acknowledgments}
AH is supported by an STFC Consolidated Grant. AH thanks Anthony Challinor, Joachim Harnois-D\'{e}raps, Alan Heavens, Peter McCullagh, Cornelius Rampf, Marcel Schmittfull, Andy Taylor and Mike Wilson for useful conversations. AM acknowledges support from a CITA National Fellowship and from NSERC. The simulations used for this project were enabled in part by support from \href{http://www.westgrid.ca}{WestGrid} and \href{http://www.computecanada.ca}{Compute Canada - Calcul Canada}.

\bibliographystyle{mn2e_fix}
\bibliography{references}

\begin{thebibliography}{48}
\expandafter\ifx\csname natexlab\endcsname\relax\def\natexlab#1{#1}\fi

\bibitem[{{Alsing} {et~al}\mbox{.}(2016){Alsing}, {Heavens}, {Jaffe},
  {Kiessling}, {Wandelt}, \& {Hoffmann}}]{2016MNRAS.455.4452A}
{Alsing} J., {Heavens} A., {Jaffe} A.~H., {Kiessling} A., {Wandelt} B.,
  {Hoffmann} T., 2016, \mnras, 455, 4452

\bibitem[{{Bernardeau} {et~al}\mbox{.}(2015){Bernardeau}, {Codis}, \&
  {Pichon}}]{2015MNRAS.449L.105B}
{Bernardeau} F., {Codis} S., {Pichon} C., 2015, \mnras, 449, L105

\bibitem[{{Bernardeau} {et~al}\mbox{.}(2002){Bernardeau}, {Colombi},
  {Gazta{\~n}aga}, \& {Scoccimarro}}]{2002PhR...367....1B}
{Bernardeau} F., {Colombi} S., {Gazta{\~n}aga} E., {Scoccimarro} R., 2002,
  \physrep, 367, 1

\bibitem[{{Bull} {et~al}\mbox{.}(2015){Bull}, {Ferreira}, {Patel}, \&
  {Santos}}]{2015ApJ...803...21B}
{Bull} P., {Ferreira} P.~G., {Patel} P., {Santos} M.~G., 2015, \apj, 803, 21

\bibitem[{{Carlson} {et~al}\mbox{.}(2009){Carlson}, {White}, \&
  {Padmanabhan}}]{2009PhRvD..80d3531C}
{Carlson} J., {White} M., {Padmanabhan} N., 2009, \prd, 80, 043531

\bibitem[{{Carron}(2011)}]{2011ApJ...738...86C}
{Carron} J., 2011, \apj, 738, 86

\bibitem[{{Carron} \& {Szapudi}(2013)}]{2013MNRAS.434.2961C}
{Carron} J., {Szapudi} I., 2013, \mnras, 434, 2961

\bibitem[{{Clerkin} {et~al}\mbox{.}(2017){Clerkin}, {Kirk}, {Manera}, {Lahav},
  {Abdalla}, {Amara}, {Bacon}, {Chang}, {Gazta{\~n}aga}, {Hawken}, {Jain},
  {Joachimi}, {Vikram}, {Abbott}, {Allam}, {Armstrong}, {Benoit-L{\'e}vy},
  {Bernstein}, {Bernstein}, {Bertin}, {Brooks}, {Burke}, {Rosell}, {Carrasco
  Kind}, {Crocce}, {Cunha}, {D'Andrea}, {da Costa}, {Desai}, {Diehl},
  {Dietrich}, {Eifler}, {Evrard}, {Flaugher}, {Fosalba}, {Frieman}, {Gerdes},
  {Gruen}, {Gruendl}, {Gutierrez}, {Honscheid}, {James}, {Kent}, {Kuehn},
  {Kuropatkin}, {Lima}, {Melchior}, {Miquel}, {Nord}, {Plazas}, {Romer},
  {Roodman}, {Sanchez}, {Schubnell}, {Sevilla-Noarbe}, {Smith},
  {Soares-Santos}, {Sobreira}, {Suchyta}, {Swanson}, {Tarle}, \&
  {Walker}}]{2017MNRAS.466.1444C}
{Clerkin} L. {et~al.}, 2017, \mnras, 466, 1444

\bibitem[{{Cohn} {et~al}\mbox{.}(2016){Cohn}, {White}, {Chang}, {Holder},
  {Padmanabhan}, \& {Dor{\'e}}}]{2016MNRAS.457.2068C}
{Cohn} J.~D., {White} M., {Chang} T.-C., {Holder} G., {Padmanabhan} N.,
  {Dor{\'e}} O., 2016, \mnras, 457, 2068

\bibitem[{{Coles} \& {Jones}(1991)}]{1991MNRAS.248....1C}
{Coles} P., {Jones} B., 1991, \mnras, 248, 1

\bibitem[{{Colombi}(1994)}]{1994ApJ...435..536C}
{Colombi} S., 1994, \apj, 435, 536

\bibitem[{{Crocce} \& {Scoccimarro}(2006)}]{2006PhRvD..73f3519C}
{Crocce} M., {Scoccimarro} R., 2006, \prd, 73, 063519

\bibitem[{{Crocce} \& {Scoccimarro}(2008)}]{2008PhRvD..77b3533C}
{Crocce} M., {Scoccimarro} R., 2008, \prd, 77, 023533

\bibitem[{{Eisenstein} {et~al}\mbox{.}(2007{\natexlab{a}}){Eisenstein}, {Seo},
  {Sirko}, \& {Spergel}}]{2007ApJ...664..675E}
{Eisenstein} D.~J., {Seo} H.-J., {Sirko} E., {Spergel} D.~N.,
  2007{\natexlab{a}}, \apj, 664, 675

\bibitem[{{Eisenstein} {et~al}\mbox{.}(2007{\natexlab{b}}){Eisenstein}, {Seo},
  \& {White}}]{2007ApJ...664..660E}
{Eisenstein} D.~J., {Seo} H.-J., {White} M., 2007{\natexlab{b}}, \apj, 664, 660

\bibitem[{{Harnois-D{\'e}raps} {et~al}\mbox{.}(2013){Harnois-D{\'e}raps}, {Yu},
  {Zhang}, \& {Pen}}]{2013MNRAS.436..759H}
{Harnois-D{\'e}raps} J., {Yu} H.-R., {Zhang} T.-J., {Pen} U.-L., 2013, \mnras,
  436, 759

\bibitem[{{Joachimi} {et~al}\mbox{.}(2011){Joachimi}, {Taylor}, \&
  {Kiessling}}]{2011MNRAS.418..145J}
{Joachimi} B., {Taylor} A.~N., {Kiessling} A., 2011, \mnras, 418, 145

\bibitem[{{Kaiser}(1987)}]{1987MNRAS.227....1K}
{Kaiser} N., 1987, \mnras, 227, 1

\bibitem[{{Kayo} {et~al}\mbox{.}(2001){Kayo}, {Taruya}, \&
  {Suto}}]{2001ApJ...561...22K}
{Kayo} I., {Taruya} A., {Suto} Y., 2001, \apj, 561, 22

\bibitem[{{Kitaura} \& {Angulo}(2012)}]{2012MNRAS.425.2443K}
{Kitaura} F.-S., {Angulo} R.~E., 2012, \mnras, 425, 2443

\bibitem[{{Larsen} {et~al}\mbox{.}(2016){Larsen}, {Challinor}, {Sherwin}, \&
  {Mak}}]{2016PhRvL.117o1102L}
{Larsen} P., {Challinor} A., {Sherwin} B.~D., {Mak} D., 2016, Physical Review
  Letters, 117, 151102

\bibitem[{Lewis {et~al}\mbox{.}(2000)Lewis, Challinor, \& Lasenby}]{Lewis2000}
Lewis A., Challinor A., Lasenby A., 2000, \apj, 538, 473

\bibitem[{{Llinares} \& {McCullagh}(2017)}]{2017arXiv170402960L}
{Llinares} C., {McCullagh} N., 2017, ArXiv e-prints 1704.02960

\bibitem[{{Matsubara}(2007)}]{2007ApJS..170....1M}
{Matsubara} T., 2007, \apjs, 170, 1

\bibitem[{{McCullagh} {et~al}\mbox{.}(2013){McCullagh}, {Neyrinck}, {Szapudi},
  \& {Szalay}}]{2013ApJ...763L..14M}
{McCullagh} N., {Neyrinck} M.~C., {Szapudi} I., {Szalay} A.~S., 2013, \apjl,
  763, L14

\bibitem[{{McCullagh}(1987)}]{McCullagh}
{McCullagh} P., 1987, {Tensor Methods in Statistics}. Chapman and Hall

\bibitem[{{Monaco} \& {Efstathiou}(1999)}]{1999MNRAS.308..763M}
{Monaco} P., {Efstathiou} G., 1999, \mnras, 308, 763

\bibitem[{{Newburgh} {et~al}\mbox{.}(2014){Newburgh}, {Addison}, {Amiri},
  {Bandura}, {Bond}, {Connor}, {Cliche}, {Davis}, {Deng}, {Denman}, {Dobbs},
  {Fandino}, {Fong}, {Gibbs}, {Gilbert}, {Griffin}, {Halpern}, {Hanna},
  {Hincks}, {Hinshaw}, {H{\"o}fer}, {Klages}, {Landecker}, {Masui}, {Parra},
  {Pen}, {Peterson}, {Recnik}, {Shaw}, {Sigurdson}, {Sitwell}, {Smecher},
  {Smegal}, {Vanderlinde}, \& {Wiebe}}]{2014SPIE.9145E..4VN}
{Newburgh} L.~B. {et~al.}, 2014, in \procspie, Vol. 9145, Ground-based and
  Airborne Telescopes V, p. 91454V

\bibitem[{{Neyrinck} {et~al}\mbox{.}(2009){Neyrinck}, {Szapudi}, \&
  {Szalay}}]{2009ApJ...698L..90N}
{Neyrinck} M.~C., {Szapudi} I., {Szalay} A.~S., 2009, \apjl, 698, L90

\bibitem[{{Nusser} \& {Dekel}(1992)}]{1992ApJ...391..443N}
{Nusser} A., {Dekel} A., 1992, \apj, 391, 443

\bibitem[{{Obuljen} {et~al}\mbox{.}(2017){Obuljen}, {Villaescusa-Navarro},
  {Castorina}, \& {Viel}}]{2017JCAP...09..012O}
{Obuljen} A., {Villaescusa-Navarro} F., {Castorina} E., {Viel} M., 2017, \jcap,
  9, 012

\bibitem[{{Peebles}(1980)}]{Peebles}
{Peebles} P. J.~E., 1980, {The Large-Scale Structure of the Universe}.
  Princeton University Press

\bibitem[{{Platz{\"o}der} \& {Buchert}(1996)}]{1996app..conf..251P}
{Platz{\"o}der} M., {Buchert} T., 1996, in Astro-Particle Physics, {Weiss} A.,
  {Raffelt} G., {Hillebrandt} W., {von Feilitzsch} F., {Buchert} T., eds., p.
  251

\bibitem[{{Schmittfull} {et~al}\mbox{.}(2017){Schmittfull}, {Baldauf}, \&
  {Zaldarriaga}}]{2017PhRvD..96b3505S}
{Schmittfull} M., {Baldauf} T., {Zaldarriaga} M., 2017, \prd, 96, 023505

\bibitem[{{Schmittfull} {et~al}\mbox{.}(2015){Schmittfull}, {Feng}, {Beutler},
  {Sherwin}, \& {Chu}}]{2015PhRvD..92l3522S}
{Schmittfull} M., {Feng} Y., {Beutler} F., {Sherwin} B., {Chu} M.~Y., 2015,
  \prd, 92, 123522

\bibitem[{{Seljak} {et~al}\mbox{.}(2017){Seljak}, {Aslanyan}, {Feng}, \&
  {Modi}}]{2017arXiv170606645S}
{Seljak} U., {Aslanyan} G., {Feng} Y., {Modi} C., 2017, ArXiv e-prints
  1706.06645

\bibitem[{{Seo} \& {Hirata}(2016)}]{2016MNRAS.456.3142S}
{Seo} H.-J., {Hirata} C.~M., 2016, \mnras, 456, 3142

\bibitem[{{Sherwin} \& {Zaldarriaga}(2012)}]{2012PhRvD..85j3523S}
{Sherwin} B.~D., {Zaldarriaga} M., 2012, \prd, 85, 103523

\bibitem[{{Simpson} {et~al}\mbox{.}(2011){Simpson}, {James}, {Heavens}, \&
  {Heymans}}]{2011PhRvL.107A1301S}
{Simpson} F., {James} J.~B., {Heavens} A.~F., {Heymans} C., 2011, Physical
  Review Letters, 107, 271301

\bibitem[{{Springel}(2005)}]{2005MNRAS.364.1105S}
{Springel} V., 2005, \mnras, 364, 1105

\bibitem[{{Sugiyama} \& {Spergel}(2014)}]{2014JCAP...02..042S}
{Sugiyama} N.~S., {Spergel} D.~N., 2014, \jcap, 2, 042

\bibitem[{{Takahashi} {et~al}\mbox{.}(2009){Takahashi}, {Yoshida}, {Takada},
  {Matsubara}, {Sugiyama}, {Kayo}, {Nishizawa}, {Nishimichi}, {Saito}, \&
  {Taruya}}]{2009ApJ...700..479T}
{Takahashi} R. {et~al.}, 2009, \apj, 700, 479

\bibitem[{{Uhlemann} {et~al}\mbox{.}(2016){Uhlemann}, {Codis}, {Pichon},
  {Bernardeau}, \& {Reimberg}}]{2016MNRAS.460.1529U}
{Uhlemann} C., {Codis} S., {Pichon} C., {Bernardeau} F., {Reimberg} P., 2016,
  \mnras, 460, 1529

\bibitem[{{Wagner} {et~al}\mbox{.}(2010){Wagner}, {Verde}, \&
  {Boubekeur}}]{2010JCAP...10..022W}
{Wagner} C., {Verde} L., {Boubekeur} L., 2010, \jcap, 10, 022

\bibitem[{{Weinberg}(1992)}]{1992MNRAS.254..315W}
{Weinberg} D.~H., 1992, \mnras, 254, 315

\bibitem[{{Wiegand} \& {Eisenstein}(2017)}]{2017MNRAS.467.3361W}
{Wiegand} A., {Eisenstein} D.~J., 2017, \mnras, 467, 3361

\bibitem[{{Zhang} {et~al}\mbox{.}(2011){Zhang}, {Yu}, {Harnois-D{\'e}raps},
  {MacDonald}, \& {Pen}}]{2011ApJ...728...35Z}
{Zhang} T.-J., {Yu} H.-R., {Harnois-D{\'e}raps} J., {MacDonald} I., {Pen}
  U.-L., 2011, \apj, 728, 35

\bibitem[{{Zhu} {et~al}\mbox{.}(2016){Zhu}, {Yu}, {Pen}, {Chen}, \&
  {Yu}}]{2016arXiv161109638Z}
{Zhu} H.-M., {Yu} Y., {Pen} U.-L., {Chen} X., {Yu} H.-R., 2016, ArXiv e-prints
  1611.09638

\end{thebibliography}

\appendix
\section{Gaussianizing transform from the Edgeworth expansion}
\label{app:edge}

In this section we derive Equation~\eqref{eq:1Dgtrans} from an Edgeworth expansion. We specialize to one dimension for simplicity, noting that the transform is unique in this scenario.

Denote by $F_X(x)$ the cumulative distribution function (c.d.f.) of $X$. The Probability Integral Transform Theorem (PITF) tells us that the quantity $Z = F_X(X)$ is uniformly distributed on the interval $[0,1]$. Denoting by $\Phi$ the c.d.f. of the standard normal distribution and using the PITF again, the quantity $Y = \Phi^{-1}[F_X(X)]$ has a standard normal distribution. Thus we can always find a transformation that Gaussianizes any non-Gaussian quantity, with the exception of distributions which cannot be completely described by their cumulants such as the lognormal~\citep{2011ApJ...738...86C}.

Now assume that the probability distribution function $P_X$ of $X$ can be written as an Edgeworth expansion
\begin{equation}
P_X = \left(1 + \frac{\kappa_3}{6}H_3 + \frac{\kappa_4}{24}H_4 + \frac{\kappa_3^2}{72}H_6 + ...\right)P_G,
\end{equation}
where $P_G(X)$ is the standard normal distribution, and the $H_n$ are Hermite polynomials. The c.d.f. is defined by $F_X(X) = \int_{-\infty}^X P_X(X') \mathrm{d}X'$, requiring us to evaluate terms of the form
\begin{align}
\int_{-\infty}^X H_n(X')P_G(X')\mathrm{d}X' &= \int_{-\infty}^X (-1)^n \frac{\mathrm{d}^n}{\mathrm{d}X^{'n}} P_G(X')\mathrm{d}X' \nonumber \\
& = (-1)^n \frac{\mathrm{d}^{n-1}}{\mathrm{d}X^{n-1}} P_G(X) \nonumber \\
& = -H_{n-1}(X)P_G(X).
\end{align}
Using the explicit form $\Phi(x) = \frac{1}{2}\left[1 + \erf \left(x/\sqrt{2}\right)\right]$ and its inverse $\Phi^{-1}(x) = \sqrt{2}\erf^{-1}(2x -1)$, we have
\begin{align}
Y(X) &= \sqrt{2} \erf^{-1} \left[\erf \left(\frac{X}{\sqrt{2}}\right) - 2\left(\frac{\kappa_3 H_2(X)}{6} + \frac{\kappa_4 H_3(X)}{24} \right. \right. \nonumber \\
& \left. \left. + \frac{\kappa_3^2 H_5(X)}{72} + ... \right)P_G(X)\right].
\end{align}
If all terms in the Edgeworth series were included, the above formula would be the exact Gaussianization transform of $X$. For weak non-Gaussianity, we can truncate the Edgeworth expansion at some order, then Taylor expand the inverse error function retaining terms of the same order to get the desired result. Doing this at $\mathcal{O}(\kappa_4)$, and using the result
\begin{equation} 
\frac{\mathrm{d}}{\mathrm{d}x} \erf^{-1}(x) = \frac{\sqrt{\pi}}{2}\exp \left\{\left[\erf^{-1}(x)\right]^2\right\},
\end{equation}
we get the result of Equation~\eqref{eq:1Dgtrans} for the Gaussianized variable.

\bsp	
\label{lastpage}
\end{document}